\documentclass[aps,prx,floatfix,twocolumn,nofootinbib,superscriptaddress]{revtex4-2}
\usepackage[dvipsnames]{xcolor}

\usepackage{graphicx,
            amsmath,
            amssymb,
            braket,
            xspace,
            lipsum, multirow}

\begin{document}

\title{Quantum error correction with the toric code}
\author{Atom Computing and Collaborators }
\date{June 2026}
\begin{abstract}
\noindent 
Quantum computing platforms based on arrays of tweezer-confined neutral atoms have recently emerged as a competitive modality thanks to a direct path toward high qubit count, rapidly advancing operation fidelities, and their ability to execute circuits with arbitrary qubit connectivity.  These features will enable the use of efficient error correction schemes with high encoding-rates, time-efficient decoding, and resource-efficient architectures based on transversal gates. With these goals in mind, recent state of the art neutral atom demonstrations focus on the transition from the use of physical qubits to error-corrected logical qubits, but to date there has been no demonstration of repeated error correction scalable to arbitrary depth. Here, we demonstrate many cycles of syndrome extraction in a toric quantum error correcting code, using mid-circuit measurement and replacement of lost qubits, including reloading of a qubit reservoir for indefinite coherent operation. We characterize the logical error rate after up to 90 cycles, showing that logical information can be preserved through multiple rounds of qubit reloading.  Comparing two distances of the code up to 8 rounds of syndrome extraction shows a lower absolute logical error rate for the larger distance code.  
\end{abstract}

\maketitle

\section{Introduction}

\begin{figure*}
    \centering
        \includegraphics[width=2.0\columnwidth]{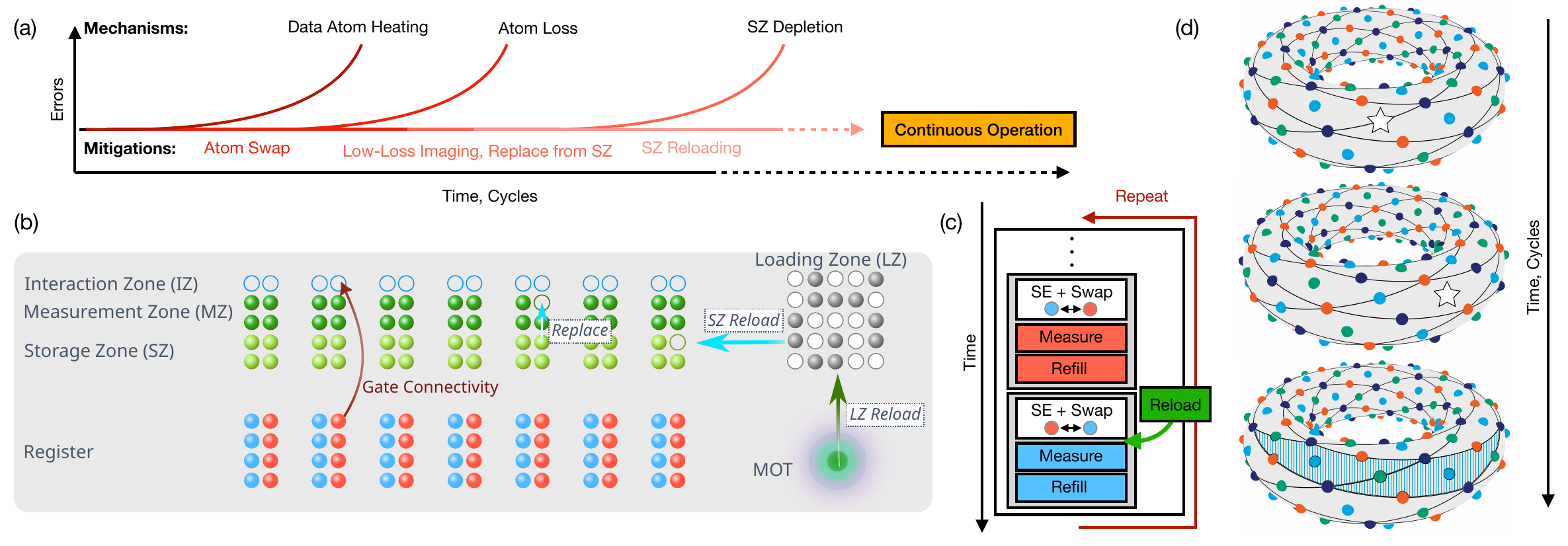}
    \caption{Architecture for continuous operation of arbitrary logical computations.  (a) Achieving continuous operation requires overcoming a series of mechanisms that limit the depth of computation that may be performed with neutral atoms.  Swapping the role of data and ancilla qubits allows all atoms to be cooled during mid-circuit measurement, mitigating the effects of heating on gate performance.  Low-loss imaging and refilling of the computational array from a local reservoir of atoms (the storage zone) extends loss-limited circuit execution, but the storage zone must be reloaded from an inexhaustible source to achieve true continuous operation.  All operations indicated must be performed without causing decoherence of existing atoms.  (b)  Our zone-based neutral-atom architecture enables arbitrary quantum computations to be performed through the use of separate static register, measurement and interaction zones.  Atoms are moved between sites using movable tweezers, enabling arbitrary connectivity.  A storage zone provides a nearby source of atoms to replace those lost in computation.  The storage zone is periodically refilled mid-circuit from a loading zone, which is in turn filled from a MOT.  (c) Continuous operation is achieved by repeating blocks consisting of several rounds of syndrome extraction, followed by reloading of the SZ.  The roles of data and ancilla qubits are swapped and missing atoms are replaced during each syndrome extraction round.  (d) A depiction of a twisted toric code across multiple cycles of syndrome extraction. $X$-type ancillas are shown on vertices, $Z$-type ancillas are shown on the faces, and data qubits are on the edges. The four colors show how the qubits swap from one cycle to the next. The star indicates a lost data qubit, that becomes a lost ancilla, which gets replaced after measurement. After completing a round of syndrome extraction, the erased qubit collapses into the wrong frame, triggering stabilizer violations shown with shaded plaquettes. 
    }
    \label{fig:1}
\end{figure*}
% motivate neutral atoms and syndrome extraction experiments
Quantum computing platforms based on arrays of tweezer-confined neutral atoms have recently emerged as a competitive modality thanks to a direct path toward high qubit count \cite{norcia2024iterative, tao2024high, manetsch2025tweezer, lin2025ai, chiu2025continuous}, rapidly advancing operation fidelities \cite{muniz2024gates, tsai_benchmarking_2024, peper2025spectroscopic, senoo2025high, evered2026high}, and their ability to execute circuits with arbitrary qubit connectivity \cite{bluvstein2022quantum, 2024AtomQEC, chinnarasu2025variational}.  These features will enable the use of efficient error correction schemes with high encoding-rates \cite{zhao2026towards, xu2025fast, yang2026spacetime, webster2026pinnacle, aasen2025}, time-efficient decoding \cite{bombin2015single, cain2025fast}, and resource-efficient architectures based on transversal gates \cite{sunami2025transversal, zhou2025low}. With these goals in mind, recent state of the art neutral atom demonstrations focus on the transition from the use of physical qubits to error-corrected logical qubits \cite{bluvstein2023logical, 2024AtomQEC, zhang2025leveraging, bluvstein2026fault, lib2026velocity}, but to date there has been no demonstration of repeated error correction scalable to arbitrary depth. Here, we demonstrate many cycles of syndrome extraction in a toric quantum error correcting code, using mid-circuit measurement and replacement of lost qubits, including reloading of a qubit reservoir for indefinite coherent operation. We characterize the logical error rate after up to 90 cycles, showing that logical information can be preserved through multiple rounds of qubit reloading.  Comparing two distances of the code up to 8 rounds of syndrome extraction shows a lower absolute logical error rate for the larger distance code.

% Recently, the error rate associated with repeated rounds of syndrome extraction has been characterized in superconducting \cite{google2025quantum}, trapped ion \cite{dasu2026computing}, and neutral-atom \cite{bluvstein2026fault} systems.  

% discuss challenges with repeated SE with neutrals

Syndrome extraction---performing measurements of a subset of qubits that reveal the presence of errors---is a key requirement for fault-tolerant quantum computation \cite{gottesman_thesis}.  
In order to be used in universal quantum computation, and to avoid a qubit number overhead that grows with circuit depth, syndrome extraction measurements must be performed in a true mid-circuit manner---the state of some qubits must be read out while coherence is preserved in others.  Recently, the error rate associated with repeated rounds of syndrome extraction using mid-circuit measurement has been characterized in superconducting \cite{google2023suppressing, google2025quantum, eickbusch2025demonstration, lacroix2025scaling, vezvaee2025surface} and trapped ion \cite{dasu2026computing} platforms, and correlations in logical observables under repeated syndrome extraction have been explored in neutral-atom \cite{bluvstein2026fault} systems.

Performing many rounds of syndrome extraction requires maintaining qubits available in a usable state, which is a challenge for atomic platforms.  Atoms are easily heated during computation, degrading the performance of future gates, and can also be lost from their traps, especially during readout.  This leads to a hierarchy of error mechanisms that become relevant at different depths, as illustrated in Fig.~\ref{fig:1}(a).  
% While the loss of a neutral atom in an optical trap is a well constrained physical error which does not necessarily proliferate throughout a processor as is possible in trapped-ion platforms \cite{coble2025correction}, 
Overcoming these challenges requires some combination of qubit reuse and replacement after measurement. Nondestructive measurement techniques can reduce the rate at which atoms need to be replaced, but true continuous operation required to execute arbitrarily deep computations requires a way to draw new atoms from an inexhaustible source such as a magneto-optical trap (MOT) without causing excessive errors in existing atomic qubits. 
% Furthermore, the circuits must be constructed in a manner that both preserves fault tolerance and ensures that each atom is measured periodically. 

% set record straight on continuous loading
The ability to form a magneto-optical (MOT) trap while maintaining coherence in computational qubits has been demonstrated using transport between spatially separated zones \cite{norcia2023midcircuit, chiu2025continuous, li2506fast}, or dual species systems \cite{singh2023mid}.  Separated zones or shelving techniques also enable iterative loading of large arrays, where new atoms can be repeatedly loaded without loss of those already present \cite{norcia2024iterative, gyger2024continuous}.  
The ability to refill tweezers either stochastically \cite{li2506fast} or in ordered arrays \cite{chiu2025continuous} from a spatially-separated MOT while maintaining coherence in existing qubits has recently been demonstrated using simple characterization circuits involving only single-qubit operations.  In this work, we integrate continuous loading, mid-circuit measurement, and qubit reuse into a fault-tolerant quantum memory circuit, showing that syndrome extraction can be performed indefinitely at constant error rate.  

% talk about error rates 

Repeated syndrome extraction forms the core of fault-tolerant schemes for quantum computation. A quantum memory experiment characterizes the performance of the logical identity operation, and is the standard test of a platform's ability to suppress the logical error rate via error correction.
While no platforms currently achieve logical error rates compatible with utility-scale computations, an observed decrease in the logical error rate as the code distance is increased is an encouraging indication that a platform may reach useful error rates upon scaling. Strictly, the threshold of a code family is an asymptotic property: the physical error rate below which the logical error rate of an infinite family of codes can be driven arbitrarily low by increasing the code distance. Recent experiments instead probe a finite-size analogue, colloquially referred to as sub-threshold behavior---a decrease in the logical error rate between two or more code sizes.
Such behavior requires that the total error rate associated with gate, idle, state preparation and measurement are sufficiently low that the added complexity of higher distance codes is offset by the increased redundancy.  Subthreshold behavior has been demonstrated in platforms using superconducting qubits \cite{google2025quantum, vezvaee2025surface}, and improvements in post-selection fraction with code distance have been shown in a trapped ion system \cite{dasu2026computing}.  These demonstrations benefit from the low error rate for physical operations currently available in those modalities, though face certain limitations: superconducting platforms are currently constrained to nearest-neighbor connectivity which restricts them to ``local" error-correcting codes, while trapped ion platforms have yet to demonstrate syndrome extraction beyond 8 cycles.  

Indications of sub-threshold behavior---evidenced by lower logical error rate for a distance-5 surface code relative to distance-3 at a fixed number of syndrome extraction cycles---have recently been reported in a neutral-atom system, though without the use of mid-circuit measurements (blocks of ancilla qubits used for syndrome extraction were retained and measured after all gate operations were performed) \cite{bluvstein2026fault}.
% Repeated rounds of a classical repetition code have been performed with mid-circuit measurement, showing logical error rates that improve with code distance, both on superconducting \cite{google2025quantum} and neutral atom platforms \cite{muniz2025repeated}.   
In this work, we characterize the logical error rates for two distances of toric codes---one using 16 data qubits and 16 ancilla qubits and the other using 32 data qubits and 32 ancilla qubits---and find lower error rates for the larger distance variant when performing 4, 6, or 8 syndrome extraction cycles. Furthermore, by implementing mid-circuit reloading, we extend syndrome extraction up to 90 cycles, and observe that the logical error rates do not increase with code distance.

Finally, we demonstrate that logical information can be maintained in our system far longer than the lifetime of any physical qubit.  Using a repetition code of distance 3 or 7 to periodically correct for errors occurring in interleaved wait-times, we show that logical information decays on a timescale of over 3 minutes for the larger distance, even when underlying qubits have lifetimes in the system on the order of 10 seconds.

% As a prerequisite for computation, quantum memory demonstrations have been performed in a variety of platforms.  In these demonstrations, repeated rounds of syndrome extraction are performed on one or more logically encoded qubits.    From the record of these measurements, the logical state of the qubit is predicted using a model of likely error sources, and compared to the logical state observed in a final measurement.  Disagreement between the predicted and observed logical state is then reported as the logical error rate.  

\section{Architecture}

We use a zone-based quantum processor \cite{bluvstein2022quantum}, with qubits defined by the ground nuclear-spin states of \textsuperscript{171}Yb atoms confined within arrays of optical tweezers.  The functional zones are shown in Fig.~\ref{fig:1}(b), and a detailed description of the system can be found in \cite{2024AtomQEC, muniz2025repeated} and references therein.  
% Our zone-based quantum processor \cite{2024AtomQEC, muniz2025repeated} hosting \textsuperscript{171}Yb nuclear spin qubits in its register, is illustrated in Fig. 1XX, with more detail described in Ref XX. 
% This architecture is further optimized by swapping the positions of the Storage Zone (SZ) and Measurement Zone (MZ) to prevent imperfect state preparation after MCM from affecting two-qubit gates in the Interaction Zone (IZ) .  
This platform enables execution of arbitrary universal quantum computation, with single-qubit gates performed in a register zone that is currently configured to support 128 qubits, two-qubit controlled-$Z$ ($CZ$) gates performed in an interaction zone (IZ), and arbitrary connectivity provided by reconfigurable parallel movement of atoms from the register to IZ.  

Beyond gates, this platform enables measurement with qubit reset, reuse, and replacement (MCM, described in  \cite{muniz2024gates}), alongside mid-circuit qubit reloading.  MCM is performed by transferring atoms in a measurement zone (MZ) into a co-located cavity-enhanced 3D optical lattice \cite{norcia2024iterative} and performing non-destructive, state-selective imaging (using techniques described in \cite{norcia2023midcircuit}), as well as state reset and cooling.  Atoms identified as missing are replaced from a near-by storage zone (SZ).  

In this work, we add the ability to reload the SZ mid-circuit from a MOT located 30~cm below the tweezer array. At pre-defined points in a computation, atoms are transported from the MOT using a moving optical lattice, and extracted into a loading zone (LZ) where light-assisted collisions (LACs) are performed.  As in our previous work \cite{norcia2024iterative}, extraction is performed without the application of cooling light.  During a subsequent MCM, the locations of atoms are identified in the LZ, MZ and SZ, and atoms from the LZ are used to refill both the MZ and SZ before circuit execution resumes see Fig.~\ref{fig:1}(c).  MOT loading and lattice transport are performed in parallel with other operations, so the only time overhead to reload comes from the LACs and the time required to move atoms from the LZ to SZ.  Details of the reloading sequence can be found in the Supplementary Material.  

% mid-circuit measurement (MCM), and continuous operation via atom reset and reload.  Crucially, continuous operation is sustained by a concurrent atom supply mechanism: atom loss identified during MCM (in MZ) is replaced by atoms from SZ, while a fresh batch is simultaneously prepared in the Loading Zone (LZ) from a magneto-optic trap (MOT). This continuous reloading and subsequent refilling of the SZ from LZ ensure a sustained qubit supply, allowing circuits to run uninterrupted.

% describeFig. 1c in some way

% \section{Toric Code Memory Experiments}
% \section{Extending Toric Code Memory with Mid-circuit measurement via Qubit Reuse and Replacement}
% \section{Co-Designing the Syndrome Extraction and Replenishment Framework}
% \section{Co-Design of a Fault-Tolerant Quantum Memory via Real-Time Qubit Reuse and Replacement}
\section{Loss-tolerant twisted toric code memory}
\begin{figure*}
    \centering
        \includegraphics[width=2.0\columnwidth]{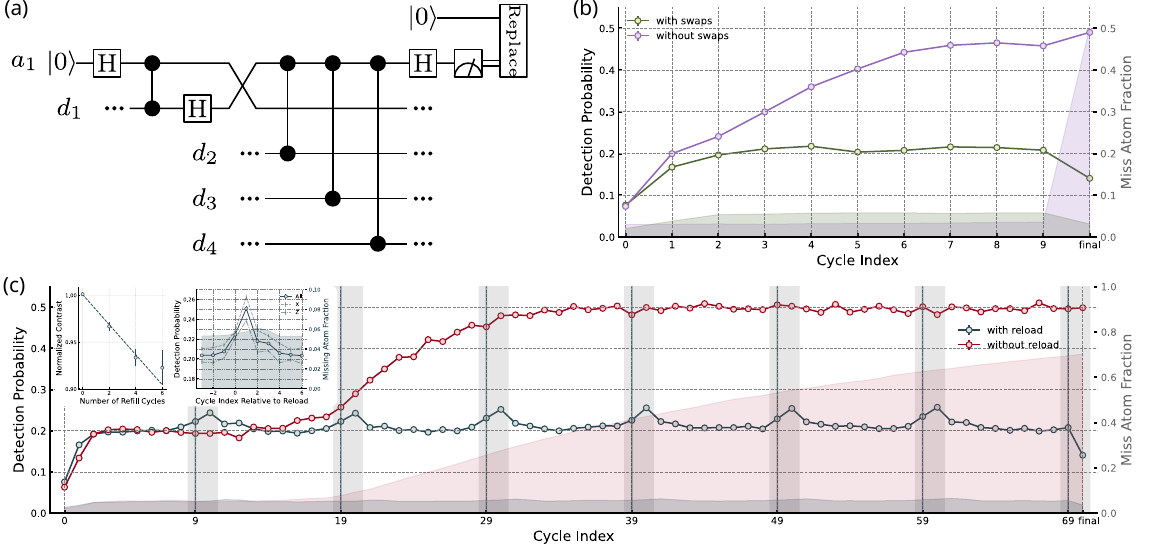}
    \caption{Toric code memory using qubit reset, reuse, role-swap, and replenishment. (a) A single $Z$-type syndrome extraction, with a compiled swap operation. An incoming data qubit interacts with a fresh qubit via a $\text{CZ}$ gate, after which their roles have been exchanged. After mid-circuit measurement, the circuit branches conditionally; a detected atom loss event triggers the replacement of the vacant ancilla site with a fresh qubit from the storage zone (SZ). (b) Stabilizer detection probability under ancilla-data swaps. Performances of a $\det 8$ Toric code over 10 consecutive syndrome extraction cycles with and without role swapping are shown in green and purple respectively. Error bars represent the standard error of the mean (SEM) and are smaller than the marker size. Shaded regions show the corresponding missing atom fractions per cycle. The final loss value of near 50\% without swaps indicates loss of nearly all data qubits.  (c) Long-term toric code memory with and without SZ refilling. A $\det 16$ toric code detection probability over 70 consecutive syndrome extraction cycles with SZ reloading every 10 cycles (olive) is compared to the no reloading baseline (red). Shaded regions track the missing atom fraction. Left inset shows the normalized MCM contrast over consecutive refilling cycles; right inset shows the cycle-averaged detection probability and missing atom fraction relative to the 10-cycle refill interval.  
    }
    \label{fig:2}
\end{figure*}
% some toric code description
% some detector frequency description, mention mcm here too
The toric code is a foundational topological quantum error-correcting code that stores information on a two-dimensional periodic lattice (represented in Fig.~\ref{fig:1}(d)), making its encoded information intrinsically robust against local perturbations \cite{kitaev2003,dennis2002}. The toric code and the related surface code are among the most well-studied quantum error correcting codes and exhibit many favorable properties for practical fault-tolerant quantum computation \cite{duclos2010,horsman2012, fowler2012}. 

It has been theoretically shown that twisting the torus on which the toric code is defined modifies its geometry while preserving the local stabilizer structure, giving access to a family of code variants with reduced qubit overhead and tunable asymmetric distances \cite{hastings2021fiber,aasen2025}. As in the standard construction, the code is defined on the two-dimensional square lattice associating qubits with edges, $X$-checks with vertices, and $Z$-checks with faces. Following Ref.~\cite{aasen2025}, we specify a twisted torus via a $2\times 2$ integer matrix $M$ in Hermite normal form. A non-zero upper right entry of $M$ describes a shear to the square lattice, introducing the twist. See Supplementary Material for a more detailed description. 

We performed a search over twisted toric codes for small instances compatible with our array geometry and movement constraints that support favorable circuit-level logical error rates.
% Following Ref.~\cite{aasen2025} we specify these codes via integer matrices in Hermite normal form:
We focus on two codes specified by:
\[
M_{\det 8} = \begin{bmatrix}
    1 & 2 \\ 0 & 8
\end{bmatrix},\ 
M_{\det 16} = \begin{bmatrix}
    1 & 3 \\ 0 & 16
\end{bmatrix}.
\]
% Any such code requires block length $n = 2\det M$ and supports two logical qubits; both code instances studied here exhibit asymmetric distances.
These codes support two logical qubits, but by treating one of the logical qubits as a gauge qubit we obtain subsystem codes with parameters 
\begin{align*}
\det 8:\ &[[16,\, 1,\, (4,\, 3)]]\\
\det 16:\ &[[32,\, 1,\, (6,\, 4)]]
\end{align*}
where we indicate the asymmetric distances $(d_X,\,d_Z)$. Note that the gauge qubit interchanges the $X$- and $Z$-distances. We refer to the logical qubit operators as $\bar{X}_1$ and $\bar{Z}_1$ and the gauge qubit operators as $\bar{X}_\mathrm{gauge}$ and $\bar{Z}_\mathrm{gauge}$. 
The lattice has a spatial self-duality which maps $\bar{Z}_1 \leftrightarrow \bar{X}_\mathrm{gauge}$ and $\bar{Z}_{\mathrm{gauge}} \leftrightarrow \bar{X}_1$. In a single $Z$-type memory experiment, we measure both $\bar{Z}_1$ and $\bar{Z}_\mathrm{gauge}$. By the lattice self-duality, the latter is equivalent to an $X$-type memory experiment for the logical qubit. Thus the same experimental sequence gives access to both the $Z$- and $X$-type memory performance of the logical qubit.

%Moreover, the spatial self-duality acts on syndrome extraction circuits by mapping an $X$-type memory experiment for the logical qubit onto a $Z$-type memory experiment for the gauge qubit.  This symmetry lets us probe both the $X$- and $Z$-type logical observables of the remaining logical qubit in a single $Z$-type memory experiment. As a result we can generically discuss the two observables of the code as interchangable.

The syndrome-extraction circuit (see Supplementary Material for a diagram) interleaves $X$- and $Z$-type stabilizers to avoid hook errors so that our circuits preserve code distance \cite{dennis2002}. A distance $d_{X/Z}$ guarantees the ability to correct $X/Z$ errors of weight $t_{X/Z}=\lfloor (d_{X/Z}-1)/2 \rfloor$. We have $t_X=1$ for the $\det 8$ code and $t_X=2$ for the $\det 16$ code. Both codes have $t_Z = 1$ and therefore offer similar protection to the $X$-type observable against Pauli noise, while $\det 16$ offers stronger protection to the $Z$-type observable. On the other hand, since heralded erasures are correctable up to weight $d-1$, the larger $d_Z$ and $d_X$ of the $\det 16$ code affords both observables strictly better protection against atom loss.

In order to mitigate the effects of atom heating and loss during repeated cycles of syndrome extraction, we incorporate role-swapping of data and ancilla qubits, as well as reset and replenishment of ancillas after measurement. By compiling periodic swap operations into the syndrome extraction procedure \cite{muniz2025repeated,chow2024leakage,liu2026_envelope}, as depicted in Fig.~\ref{fig:2}(a), encoded information is preserved while allowing each atom to be reset after a limited number of operations---a typical atom participates in two syndrome extraction cycles before being measured, reset, and if necessary, replaced. 

% We perform repeated cycles of syndrome extraction with mid-circuit measurements for $X$ and $Z$ stabilizers.
% Atom heating and losses are among the key challenges in performing long quantum error correction circuits on neutral atom based quantum processors. We mitigate these errors by incorporating qubit role-swap and replenishment throughout cycles of syndrome extractions. In a \emph{role-swapping} protocol, the encoded information is preserved by compiling periodic swap operations into the syndrome extraction procedure, effectively exchanging the roles of data and ancilla qubits  \cite{muniz2025repeated,chow2024leakage,liu2026_envelope} as depicted in Fig.~\ref{fig:2}(a). At the beginning of a syndrome extraction cycle, a fresh qubit $a_1$ interacts with a data qubit $d_1$ in a particular stabilizer, transferring the information from $d_1$ to $a_1$. After this interaction, the former data qubit serves as an ancilla, interacting with the remaining data qubits $d_2, d_3,\ldots$ in the check. Similarly, the former ancilla becomes a data qubit and interacts with ancilla qubits in neighboring checks. At the end of the round, the qubits serving as ancillas are measured, cooled, and---if necessary---replaced, in order to begin the process again. Thus a typical atom participates in two syndrome extraction cycles before being measured, with its label following the trajectory $a\rightarrow d\rightarrow a$.
% This approach bounds the number of operations any qubit is subjected to before measurement and reinitialization. 

Detectors represent inconsistent correlations between syndromes of consecutive cycles \cite{fowler2014scalable}, and the detection probability $P_d$ can be used to characterize the temporal stability of physical error rates throughout the execution of the circuit \cite{google2025quantum,hesner2024}.  Our imaging method used for qubit measurement independently identifies qubit state and loss, enabling us to perform delayed erasure conversion \cite{Wu2022erasure, zhang2025leveraging}. We handle loss events by randomly assigning a state prior to computing detection probabilities.
% To compute detection probabilities in which a lost atom was expected to contribute a bit value, we randomly assign values prior to computing detection probabilities. In an otherwise error-free circuit, a single loss before a measurement would then trigger 0 or 2 detectors with equal probability, leading to an average of 1 detection event per loss, versus weight-1 Pauli errors that contribute 2 detection events per error. The reduced sensitivity of detection probability to loss is consistent with reduced logical sensitivity to erasure error---although we note that atom loss is not immediately detected and is therefore not strictly an erasure error.

\section{arbitrarily repeatable syndrome extraction}

The combination of role-swapping, ancilla reuse and replacement, and reloading of atoms from the MOT combine to enable continuous syndrome extraction for arbitrary numbers of rounds.  In Fig.~2(b), we compare $P_d$ with or without role-swapping when operating the $\det 8$ toric code over 10 cycles of syndrome extractions.
In the static configuration with fixed data and ancilla roles, $P_d$ rises steeply due to heating and loss of data qubits. In contrast, with the swap protocol, $P_d$ remains near a constant baseline of $\approx0.2$ (the cycles indexed 0 and 1, as well as the final cycle are expected to show lower $P_d$ due to edge-effects). The elevated ancilla qubit loss observed in the role-swapping configuration (see green and purple shades inFig.~\ref{fig:2}(b)) results from the fact that a measured atom experiences 8 two-qubit gates between measurements with role-swapping, compared to 4 two-qubit gates without. This demonstration shows the effectiveness of resetting and periodically replacing atoms as enabled by the role-swapping protocol.

%While role-swapping introduces a marginal elevation in loss rates on the measured ancilla qubits due to the overhead of reconfiguration (see green and purple shades in Fig.2(b)), this localized error does not propagate into a global failure due to the temporal-isolation provided by role-swappings. \\
% some refilling reloading description

% While role-swapping ensures that operation-induced heating does not continuously accumulate on any single qubit, the sustainability of the logical memory is ultimately governed by the physical presence of the atoms themselves. 

While role-swapping and and ancilla replacement prevents the buildup of heating and loss errors on specific atoms, this process will eventually deplete the occupancy of the SZ. To mitigate this, we employ mid-circuit reloading of the SZ once every 10 cycles of syndrome extraction.  
In Fig.~2(c), we track $P_d$ for a $\det 16$ toric code over 70 cycles of syndrome extraction (the total number of cycles is currently limited by compilation overhead). Without any atom replenishment to SZ (red points, shaded region), we observe a rapid increase in $P_d$ as accumulated atom losses can no longer be addressed upon depletion of SZ, occurring here around cycle $15$. 
%Our replenishment strategy counteracts atom loss due to vacuum collisions or measurement-induced heating operating on two distinct timescales: every cycle, ancilla losses detected in the MZ following MCM are immediately replaced with fresh atoms from the SZ; every 10 cycles, we refill the SZ from the LZ—which is continuously loaded from a Magneto-Optical Trap (MOT).
%Following each mid-circuit measurement, any tweezer site identified as vacant is restored through a \textit{replace} operation, where fresh qubits are shuttled from a storage zone (SZ) into the empty sites in the measurement zone (MZ). To prevent depletion of this buffer, a background process periodically reloads the loading zone (LZ) from a Magneto-Optical Trap (MOT) and subsequently \textit{refills} the SZ from LZ. 
% The experimental impact of this replenishment strategy is illustrated in Fig. 2(c), which tracks $P_d$ for a $\det 16$toric code over 70 cycles of syndrome extraction with SZ replenishment (green) and without (red). Initially, $P_d$ remains nearly identical for both versions, as both use atom replacement from the SZ. However, upon depletion of SZ, occurring here around cycle 12, the `without refill' case exhibits a rapid increase in $P_d$ as accumulated atom losses can no longer be addressed. 
In contrast, the `with-refill' case (gray points, shaded region) maintains a steady-state $P_d$ indefinitely, albeit with periodic increase of a few $~\%$ during the refill cycles (indicated by vertical dashed lines in Fig.~2(c)).  

The elevated detector frequency can be caused by decoherence induced by the reloading process.  We estimate the magnitude of the decoherence by comparing the contrast of two Ramsey sequences: a baseline condition with MCM and a hold-time equal to the $\det 16$ syndrome extraction cycle time containing a single echo pulse, and the same sequence with a full reloading procedure included in parallel with the hold time.  The relative contrast versus the number of blocks included is shown in the inset of Fig.~\ref{fig:2}(c), and indicates an added $1.6(1)\%$ contrast loss per refilling cycle. While this is a coherent error that will accumulate differently depending on what operations are performed in parallel with reloading, it is on the right scale to explain the bumps in detection probability.  

The elevated detection probability typically persists for 3 cycles, with no significant increase ($<1\%$) in the missing atom fraction (see first inset of Fig.~\ref{fig:2}(c)). The first two cycles of elevated $P_d$ are a direct consequence of the differential nature of the detector (indicated by vertical shades in Fig.~\ref{fig:2}(c)), which compares parity outcomes between successive cycles. The observation of two additional elevated cycles indicates a brief entropy dissipation period, during which repeated mid-circuit measurement (MCM) and reset operations remove the injected noise. 

% We characterize the measurement error induced by the refilling process by measuring MCM contrast losses over multiple consecutive rounds (see second inset of Fig.2(c)); from this, we extract a $1.6(1)\%$ contrast loss per refilling round. This is on the same order of magnitude as the observed $P_d$ spikes, confirming that the transient noise is primarily refilling-induced. Despite this multi-round signature, these errors remain strictly localized in time and do not compromise the long-term stability of the logical state.
% with and without reloading comparison

\section{Logical Errors Suppression with Code Distance}

\begin{figure*}
    \centering
        \includegraphics[width=2.0\columnwidth]{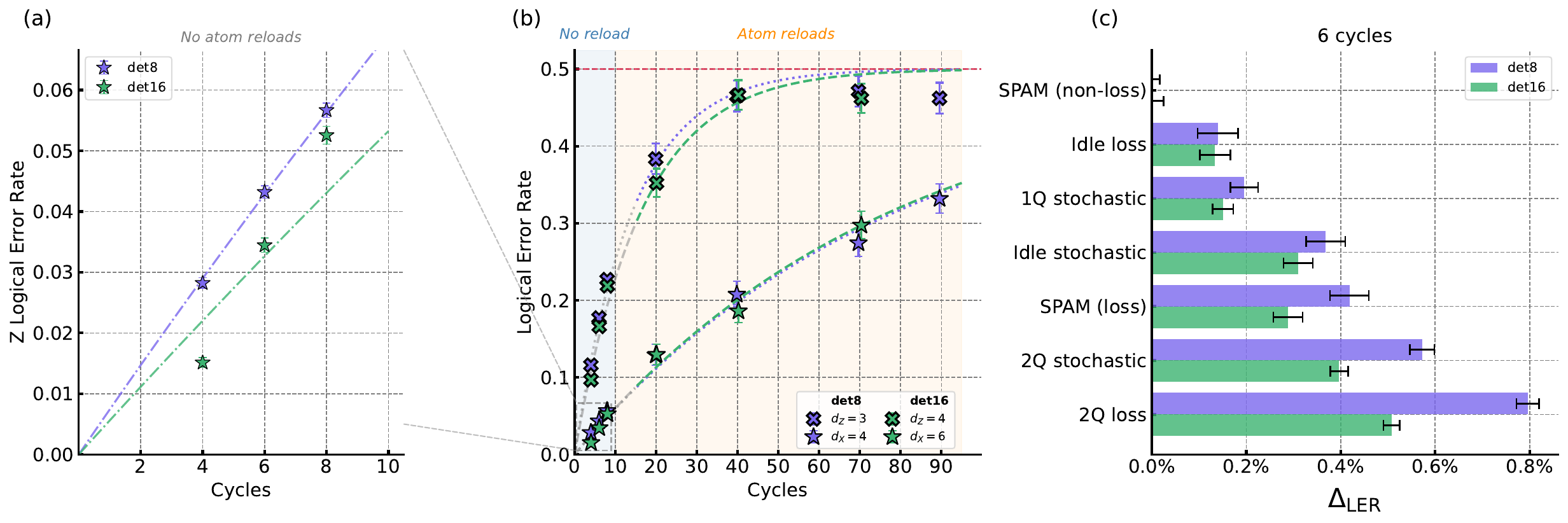}
    \caption{(a) $Z$-type logical error rates without SZ reloads averaged over 6 datasets.
    % comprising over 180 hours of experiment time. 
    The $\det 16$ code shows logical error rates at or below those of the $\det 8$ code, even through fluctuations over time within this dataset (see Supplementary Material). Lines are fits to a fixed logical error per cycle model to guide the eye, $\mathrm{LER}(c)=\frac{1}{2}(1-(1-2\epsilon_{\mathrm{cycle}})^c)$ \cite{google2025quantum,bluvstein2026fault} with parameters $\epsilon^Z_{\mathrm{cycle,det8}}=0.74\pm.01\%$ and $\epsilon^Z_{\mathrm{cycle,det16}}=0.56\pm.01\%$. The $\det 16$ data shows significant deviations from this model likely due to boundary detector effects and slow atom heating (see Supplementary Material). (b) Logical error rate per cycle for both the $Z$- and $X$-type extended to deep cycles by reloading atoms to the SZ every 10th cycle. The $X$-type observable is measured through the $Z$-type observable of the gauge qubit. Dotted purple (det8) and dashed green (det16) lines are fits including only data with reloads to the fixed logical error per cycle model with parameters $\epsilon^Z_{\mathrm{cycle,det8}}=0.63\pm.03\%$, $\epsilon^X_{\mathrm{cycle,det8}}=3.4\pm.3\%$, $\epsilon^Z_{\mathrm{cycle,det16}}=0.64\pm.04\%$ and $\epsilon^X_{\mathrm{cycle,det16}}=3.0\pm.2\%$. Atom reloading does not appear to have a significant effect on logical error rate per cycle. (c) $Z$-type error gradients for the $\det 8$ ($d_X=4$, purple) and $\det 16$ ($d_X=6$, green) codes at six cycles simulated from an independently calibrated error model (see Supplementary Material).
    }
    \label{fig:3}
\end{figure*}
We next characterize the logical error rate for the two code variants after different numbers of syndrome extraction cycles.  
%As described above, the conjugate nature of the two toric code logical qubits allows for extraction of the logical error rates for both the $Z$ and $X$ observable of one logical qubit in a single experiment sequence by measuring the $Z$ observable of both logical qubits. 
As described above, lattice self-duality allows a single $Z$-type memory experiment to access logical error rates for both the $Z$- and $X$-type logical observables of one logical qubit: the $Z$ logical observable is measured directly, while the complementary $X$ logical observable is obtained through the dual $Z$-type observable of the gauge qubit.
The final cycle uses a transversal measurement of the ancilla and data qubits to obtain the two final rounds of detectors as well as measurements of the logical observables.

% Leveraging the ability to perform circuits to arbitrary depth without a persistent increase in detection probability or atom loss, we execute a toric code memory experiment with varying numbers of cycles using the $\det 8$ and $\det 16$ codes defined in the previous section. 

% We prepare all qubits in the $Z$-basis and perform repeated cycles of syndrome extraction using role-swapping and MCMs with atom replacement. The final cycle uses a transversal measurement of the ancilla and data qubits to obtain the two final rounds of detectors as well as measurements of the logical observables. The conjugate nature of the two toric code logical qubits allows for extraction of the logical error rates for both the Z and X observable of one logical qubit in a single experiment sequence by measuring the Z observable of both logical qubits. 

The recorded bitstrings were decoded using PyMatching, a minimum-weight perfect matching decoder \cite{fowler2013optimalcomplexitycorrectioncorrelated,Higgott2025sparseblossom}. We use Stim to construct an initial detector error model \cite{Gidney2021stimfaststabilizer,Derks2025}. On a per-shot basis, we use a heuristic method to re-weight the prior in the detector-error model around atom loss events (further details are provided in the Supplementary Material). Notably, this procedure requires a significantly reduced computational overhead compared to supercheck decoders \cite{stace2009thresholds,barrett2010}.

In Fig.~\ref{fig:3}, we present the logical error rate for both codes in regimes that do and do not include mid-circuit reloading from a MOT.  All conditions involve role-swapping, MCM with reset, and replacing lost ancilla qubits from the SZ.

Memory experiments out to 8 cycles do not require reloading the storage zone (SZ). To ensure consistent machine performance for the two codes, 35 shots of the $\det 8$ codes were interleaved with 20 shots of the $\det 16$ codes across 6 separate datasets totaling approximately 180 hours of data collection ($\approx$39,000 shots per $\det 8$ point and $\approx$23,500 shots per $\det 16$ point). Fig.~\ref{fig:3}(a) shows the decoded logical error rates as a function of cycles with overlaid fits to a fixed logical error per cycle, $\epsilon_{\mathrm{cycle}}$ model \cite{google2025quantum,bluvstein2026fault} with parameters $\epsilon^Z_{\mathrm{cycle,det8}}=0.74\pm.01\%$ and $\epsilon^Z_{\mathrm{cycle,det16}}=0.56\pm.01\%$. The $\det 16$ data do not appear to follow a fixed logical error rate per cycle in this range, likely due to extended boundary effects from the high initial state preparation fidelity and increased loss with cycles due to slow atom heating (see additional data and simulations in the Supplementary Material). We note that directly converting the 4-cycle point to a logical error rate per cycle as in \cite{bluvstein2026fault} yields $\epsilon^Z_{\mathrm{cycle,det16}}=0.38\pm .02\%$ and an error suppression factor $\Lambda^Z_{\text{no reload}}=1.9\pm.1$. 
Similar behavior is observed in the $X$-observable at low cycles (Fig.~\ref{fig:3}, right), but with significantly higher logical error rates and a smaller error suppression factor (again at 4 cycles) $\Lambda^X_{\text{no reload}}=1.20\pm.03$.  Averaging over the two observables, we observe $\Lambda^{\text{Avg}}_{\text{no reload}}=1.30\pm.03$.  

While the observed decrease in logical error rate for the higher-distance code is consistent with sub-threshold behavior in this regime, several aspects of the result suggest care in extrapolating these results.
% in the $\det 16$ code than the $\det 8$ code for all distances explored, we avoid making strong claims to sub-threshold behavior for two reasons.  
First, our error probability exhibits trends versus numbers of cycles that are not well-captured by an error model that neglects a careful treatment of coherent errors and atom heating.  This observation merits caution in extrapolating per-cycle error rates from measurements performed at a limited sampling of cycle numbers, even if backed by a stochastic error model.  Second, we cannot rule out the possibility of per-operation error rates that grow with system size. While this is a somewhat generic statement about hardware, it may be more relevant for systems that take advantage of arbitrary connectivity.

% . This is a result of biasing the syndrome extraction to favor one logical qubit and the conjugate nature of the toric code logical qubits (i.e. the logical errors are biased into the second qubit's Z-observable which has the same logical error rate as the first qubit's X-observable). We note that biasing logical errors into one sector is not universally desirable for fault-tolerant quantum computing; however, this still represents a coherent quantum memory experiment as the X-observable of the second logical qubit is also preserved. In codes without conjugate logical qubits, this technique provides an avenue for reducing logical error rate at the cost of expending a logical qubit.

% , although we note that fluctuations in experiment performance over time within this dataset show deviations from this behavior (see supplementary material). With this in mind, this demonstrates logical error rates for a distance-6 quantum error-correcting code that are below or consistent with those observed at distance-4. 

% To go beyond 8 cycles, we add in reloading operations of the SZ evert 10th cycle.  The complete dataset shows suppressed logical error rates for the $\det 16$ ($d_Z=6$) code compared to the $\det 8$ ($d_Z=4$) code.  

Reloading the SZ every 10th cycle extends the memory experiment out as far as 70 cycles for the $\det 16$ code and 90 cycles for the $\det 8$ code (Fig. ~\ref{fig:3} (b); again, depth is currently limited by compilation overhead). The two codes were interleaved at 10 shots each within a single dataset spanning 67 hours ($\approx$660 shots per $\det 8$ point and $\approx$710 shots per $\det 16$ point). 
We find reasonable agreement between data with reloading and the fixed logical error rate per cycle model. The resulting fits yield $\epsilon^Z_{\mathrm{cycle,det8}}=0.63\pm.03\%$ and $\epsilon^Z_{\mathrm{cycle,det16}}=0.64\pm.04\%$. This demonstrates that over an arbitrary number of cycles, the logical error rates do not increase with larger code distance. Importantly, atom reloading does not appear to cause a significant increase in logical error rates per cycle.

We simulate errors in our system using a model that accounts for independently calibrated stochastic errors as well as loss (see Supplementary Material for a description of the simulation and input parameters).  As a reference case, we consider the $Z$-type logical error rates after six cycles here---a broader comparison between the model and observed logical error rates, detector frequencies, and loss is explored in the Supplementary Material. Our model predicts $Z$-type error rates of 1.1\% for $\det 16$ and 1.9\% for $\det 8$, somewhat lower than the average observed values of Fig.~\ref{fig:3}, but similar to the best-performing time block of the data (see Supplementary Material).  This difference may be attributed to potential coherent errors which are not captured in our model (for example, uncalibrated qubit frequency differences or drifts that are imperfectly canceled by spin-echo techniques), or to difficulties in accurately calibrating the in-situ performance of operations, where effects like atom heating may be relevant.  Incorporating additional data qubit loss in the model partially resolves this discrepancy and better captures certain features of detector frequencies and loss (see Supplementary Material).

In Fig.~\ref{fig:3}(c) we represent the simulated sensitivity of $Z$-type logical error rates after six cycles to the calibrated physical error sources by plotting the simulated reduction in the logical error rate when that error source is removed entirely.  Because logical error rates are not linear versus physical error rates, we do not expect the component sensitivities to add up to the observed logical error rate, but do expect the relative values to be representative of the marginal contribution of the associated error source.  Two features are apparent from this analysis: the 2Q gates contribute the largest single source of error, and the various channels for loss, taken together, exceed the contribution from stochastic errors.

\begin{figure*}
    \centering
        \includegraphics[width=1.0\columnwidth]{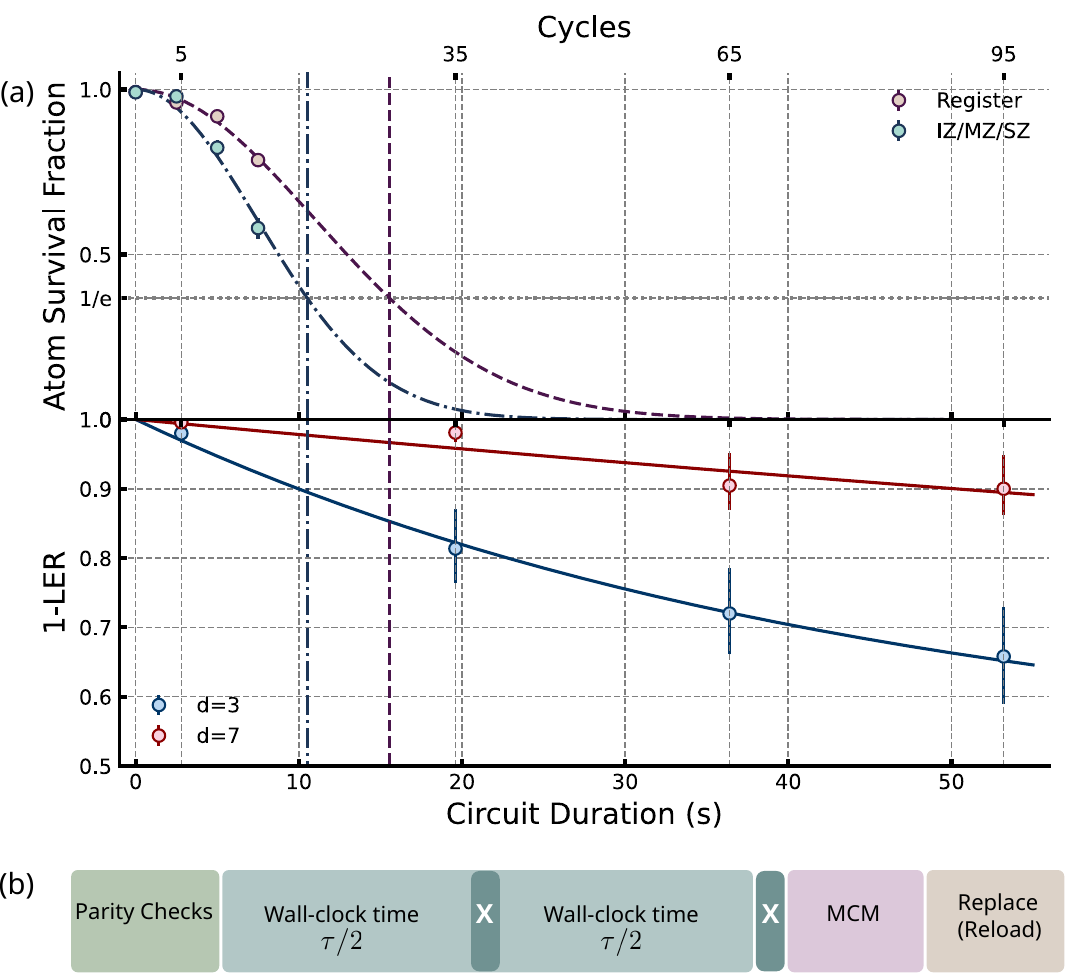}
    \caption{Sustained Quantum Memory beyond physical qubit lifetime. (a) Repetition code performance over extended operational times. The top panel shows physical atom survival fractions over time in the register (purple) and science zones (IZ, MZ, and SZ; green). The bottom panel displays logical error rates for the $d=3$ (red) and $d=7$ (blue) repetition codes, plotted against dual axes mapping both the total number of QEC cycles and cumulative circuit durations.  (b) Timing diagram of a QEC cycle extended by wall-clock idling time. Following the stabilizer parity checks, qubits idle in the register for a duration $\tau/2$, after which an echo $\pi$-pulse is applied. Following a second $\tau/2$ idling window, a final $\pi$-pulse restores the qubits to their original states, immediately followed by MCM and atom replacement, and SZ reloading every 20th cycle. 
    }
    \label{fig:4}
\end{figure*}

\section{A Robust Repetition Code Memory Surpassing Physical Qubit Decay}
% While ... (say something about the on threshold behavior from last section), the ultimate validation of our dynamic replenishment framework lies in its operational endurance. 
% Across the broader quantum error correction landscape, extending an operational logical lifetime past the limits of its underlying physical components is a vital architectural milestone and it demands highly platform-specific engineering. In superconducting architectures, the race is often against the rapid energy relaxation ($T_1 \sim 10\ \mu\text{s}$ to $100\ \mu\text{s}$), with competing methodologies focusing either on drastically reducing the QEC cycle time(cite google) or introducing hardware innovations aimed at fundamentally extending $T_1$ via bosonic cavity states (cite yale, double check accuracy). Meanwhile, trapped-ion architectures operate in a regime where physical population lifetimes are practically infinite, allowing experiments to focus entirely on extending the shorter dephasing lifetime ($T_2$) by nesting states in decoherence-free subspaces (cite quantinuum) or using autonomous, dissipative spin-to-phonon cooling (cite USTC).\\
One of the defining milestones of fault-tolerant quantum computing is the creation of a logical qubit with an operational lifetime that surpasses the decay of its physical components, which is essential for preserving quantum information indefinitely. In neutral-atom platforms, the operational upper bound is fundamentally governed by physical atom loss from the optical traps. To establish a rigorous operational milestone, we benchmark our encoded logical memory directly against this physical atom survival baseline. We characterize idle atom survival independently within both the science zones (IZ, MZ and SZ) and register using a gaussian decay fit, extracting baseline atom $1/e$ lifetimes to be $10.5(4)$ s and $15.6(5)$ s, respectively (Fig.~\ref{fig:4}(a) top).
We measure the logical memory of distance 3 ($d=3$) and distance 7 ($d=7$) phase-sensitive repetition codes \cite{muniz2025repeated} up to 95 QEC cycles with MCM and atom replacement every cycle, complemented by an SZ refill routine every 20 cycles (Fig.~\ref{fig:4}(a) bottom). We intentionally introduce extra wall-clock idling time ($\tau = 0.5$s) into each individual cycle (see Fig.~\ref{fig:4}(b)), artificially stretching the total circuit duration to expose the encoded logical states to maximum environmental error. The value $\tau=0.5$~s was chosen as a practical operating point, balancing increased idle error against logical error suppression per unit time; it is not intended to be an optimized delay time. Remarkably, the logical error rates for both $d=3$ and $d=7$ repetition codes remain strictly below the completely random $0.5$ limit well beyond the physical lifetime of atoms held in tweezers. To quantify this behavior, we plot $1 - \text{LER}$ and characterize the characteristic logical lifetime by fitting the decay profile to $0.5 + 0.5\exp(-t/\tau_{\text{L}})$, from which we extract logical lifetimes of $\tau_{\text{L}} = 44(1)$ s for the $d=3$ and $\tau_{\text{L}} = 225(33)$ s for $d=7$ repetition codes respectively, for the specific idle times chosen. This persistent error suppression over macroscopic operational duration demonstrates a sustained logical memory that outlasts the physical presence of individual atoms.

\section{Outlook}
In this work we have demonstrated error suppression in a toric code quantum memory that integrates all of the physical components necessary for continuous operation of an error-corrected neutral atom quantum processor for the first time. Critically, we have demonstrated a syndrome extraction scheme that frequently swaps the identity of data and ancilla qubits such that lost qubits are quickly identified as an erasure errors and replaced by freshly initialized qubits. This scheme is generalizable to other QEC codes that will be used in utility-scale quantum computers. To further demonstrate the viability of this platform, we operated a repetition code logical memory maintaining encoded information up to one minute, well beyond the lifetime of the atomic qubits. The reduction in logical error rate with increased code distance is consistent with sub-threshold performance of the quantum processor.
% despite the presence of physical errors incurred during continuous atom reloading. 
Making use of the flexible connectivity of neutral atom qubits, this is the first demonstration of continuous syndrome extraction in a code that requires non-local connectivity inaccessible in a 2D planar geometry and enables practical utility-scale codes.  

% represents a key step towards the use of block codes with high encoding rates.  

% While utility-scale architectures are likely to use block codes with higher encoding rates [cite architecture papers] the toric code is a prototypical example of a non-2D planar code encoding multiple logical qubits per code block. 

Moving from these demonstrations to useful, utility-scale quantum computing requires advancement on multiple fronts. We anticipate that improved two-qubit gate performance (particularly loss) can be achieved through the use of faster gate schemes that are less sensitive to atomic temperature, which in turn will enable faster atomic movement at acceptable heating levels and lower associated idle errors as well as cycle times.  The system size may be increased to utilize more advanced error correction codes with larger distance and logical operations (Clifford and non-Clifford) with efficient space-time cost.  This will require both increased laser power and the ability to maintain flexible parallelization of atom movement---for example with multiple independent sets of mobile tweezers.  Further, while the moderate system size and low-loss operations demonstrated in this work mean that steady-state operation can be achieved with a modest flux of atoms, increased system size and decreased logical cycle times may benefit from performing either multiple extractions of atoms from the transport lattice per loading cycle \cite{li2506fast, chiu2025continuous}.

% lower physical error rates well below threshold, higher-distance error correction codes to achieve utility-scale logical error rates, more advanced error correction codes to provide memory and logical operations (Clifford and non-Clifford) with efficient space-time cost. 

\onecolumngrid

\vspace{1em}
\begin{flushleft}
{\small Atom Computing and Collaborators}

\bigskip
{\small
\renewcommand{\author}[2]{#1$^\textrm{\scriptsize #2}$}
\renewcommand{\affiliation}[2]{$^\textrm{\scriptsize #1}$ #2 \\}
\newcommand{\xprimary}{\affiliation{*}{These authors contributed to data taking, analysis and simulation for this work.  }}

\newcommand{\xAtom}{\affiliation{1}{Atom Computing, Inc.}}

\newcommand{\xMSFT}{\affiliation{2}{Microsoft Quantum}}

\newcommand{\xStanford}{\affiliation{3}{Department of Physics and Department of Applied Physics, Stanford University}}

\newcommand{\xMontana}{\affiliation{4}{Department of Physics, Montana State University}}

\newcommand{\primary}{*}
\newcommand{\Atom}{1}
\newcommand{\MSFT}{2}
\newcommand{\Stanford}{3}
\newcommand{\Montana}{4}

\author{David Aasen}{\MSFT},
\author{Alexander Aeppli}{\Atom},
\author{Stephen Armstrong}{\Atom},
\author{Sambit Banerjee}{\Atom},
\author{Katrina Barnes}{\Atom},
\author{Ethan Becker}{\Atom},
\author{Juan M. Bello-Rivas}{\MSFT},
\author{Trent Bjorkman}{\Atom},
\author{Benjamin J. Bloom}{\Atom},
\author{Ian Bloomfield}{\Atom},
% ...

\bigskip
\xprimary
\xAtom
\xMSFT
\xStanford
\xMontana
}
\end{flushleft}

\clearpage

\bibliography{bib}
% \addbibresource{bib}
% --- END OF MAIN MANUSCRIPT TEXT ---

\clearpage 
\onecolumngrid 

\begin{center}
    \textbf{\large Supplemental Material}

    \vspace{12pt}
    \small Atom Computing and Collaborators\\
    \textit{Dated: \today}
\end{center}

\vspace{20pt}

% Setup Supplementary numbering
\setcounter{equation}{0} \renewcommand{\theequation}{S\arabic{equation}}
\setcounter{table}{0}    \renewcommand{\thetable}{S\arabic{table}}
\setcounter{figure}{0}   \renewcommand{\thefigure}{S\arabic{figure}}

\setcounter{section}{0}  
\renewcommand{\thesection}{S\Roman{section}} 

% This resets the citation numbering back to 1 and adds the "S"
\setcounter{enumiv}{0}              
\renewcommand{\citenumfont}[1]{S#1} 
\renewcommand{\bibnumfmt}[1]{[S#1]} 

\twocolumngrid 

% Reset bibliography counter for supplement
\setcounter{enumiv}{0}
\makeatletter
\renewcommand{\@biblabel}[1]{[S#1]}
\makeatother
%   \documentclass[aps,prx,floatfix,twocolumn,nofootinbib,superscriptaddress]{revtex4-2}
% \usepackage[utf8]{inputenc}
% \usepackage{amsmath,amssymb} % Helpful for supplementary equations
% \usepackage{graphicx}
% \usepackage{xr}
% \usepackage{booktabs}
% \usepackage{multirow}
% \externaldocument{main}

% % Setup Supplementary numbering for equations and tables only
% \setcounter{equation}{0}
% \renewcommand{\theequation}{S\arabic{equation}}
% \setcounter{table}{0}
% \renewcommand{\thetable}{S\arabic{table}}

% \title{Supplementary Information}
% \author{Atom Computing, Inc.}
% \affiliation{Atom Computing, Inc.}
% \date{\today}

% \begin{document}
% \maketitle

\section{SZ Reloading Sequence}

Extended Data Fig.~\ref{Extended Data Fig:ext_reload_time} illustrates the sequence for atom reload operations. Detailed physical implementations of each operation were described in previous work~\cite{muniz2025repeated}. The order of operations is adapted for parallel circuit execution, while their effects on the circuits are discussed in the main text. 

The MOT and transport stage happens in parallel with the syndrome extraction (SE) circuit. Background loading of the MOT contributes an additional $0.014(6)$~$1/\mathrm{s}$ to the overall decoherence rate.  Before rearrangement moves atoms to the MZ for MCM, but after the ancilla atoms are brought to the measurement basis, a light-assisted collision (LAC) block prepares the LZ with a $50\%$ average filling. The LAC beams are local to the LZ, IZ, MZ, and SZ. To echo out the residual light shift from the LAC beams to finite extinction in the RZ, a composite global $\pi$ pulse is added between the two LAC blocks. 

%An extra $\pi$ pulse is applied after the second LAC to flip the RZ atom state back. 

% The composite $\pi$ pulses, including those used later in the MCM block, are realized via a global beam illuminating the entire array. This single beam is driven with two RF tones separated in frequency by the qubit transition frequency. The polarization is carefully tuned to balance the light shifts on the two qubit states. Each component pulse takes a rounded, flat-top pulse shape with a $\pi$-pulse duration of $\sim 65~\mu\text{s}$. The composite pulse is a 5-pulse Knill type with suppression of second order off-resonance errors. Refer to TABLE I. in~\cite{Jones2013Designing} for their Figure 5(b). 

% The composite $\pi$ pulses, including those deployed later within the MCM block, are realized via a global beam illuminating the entire atomic array. This single beam is driven with two RF tones separated in frequency by the qubit transition frequency, with the polarization carefully tuned to balance the light shifts on both qubit states. Each component pulse features a rounded, flat-top profile with a single $\pi$-pulse duration of $\sim 65~\mu\text{s}$. To suppress second-order off-resonance errors, the composite sequence utilizes a five-pulse Knill-type configuration, implementing the specific phase sequence detailed in Table I (specifically corresponding to Fig. 5b) of Ref.~\cite{Jones2013Designing}.

Inside the MCM block, a similar echo strategy is deployed between two MCM images.  The first image is state-selective and the second is state-insensitive (optical pumping is applied) for atom-loss detection. Following the MCM block, all atoms in the LZ, MZ, and SZ are cooled and reinitialized. Subsequently, our software service uses the MCM images to determine occupancy and calculate the rearrangement moves required to refill both the SZ and MZ from the LZ, prioritizing the MZ. The overall duration of the refilling process depends on the number of sites to fill and their specific locations; on average, it takes approximately $90~\text{ms}$, which includes the software processing time and a sequence buffer time of $\sim 20~\text{ms}$.

To mitigate the effects of static qubit frequency differences or drifts, we deploy a dynamical decoupling echo at the midpoint of each SE circuit, when all active data and ancilla qubits occupy the static register sites. 
% An example circuit encompassing this protocol across 4 cycles (excluding atom rearrangement routines) is available as Supplementary Data 1 and 2 (\texttt{Supplementary\_Data\_1-det8-4cycles.stim} and \texttt{Supplementary\_Data\_2-det16-4cycles.stim}, respectively) in the native Stim file format.

\begingroup
\renewcommand{\figurename}{Extended Data Fig.}
\begin{figure*}
    \centering
    \includegraphics[width=2.0\columnwidth]{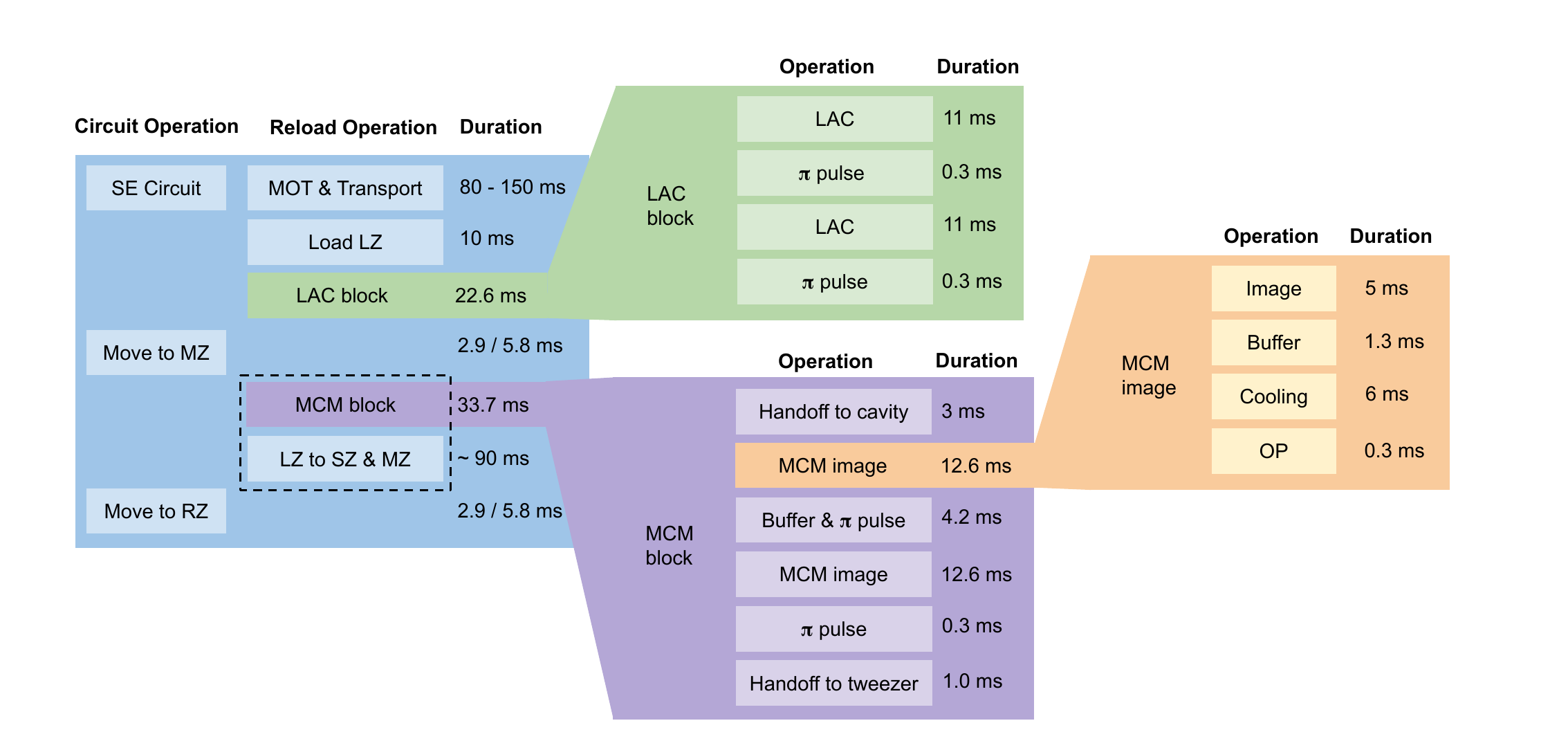}
    \caption{Timing for a SE cycle with SZ reloading. The duration for MOT and transport, $\approx 60$ ms,  is adjusted to be shorter than the length of the SE circuit, which is roughly 80 ms for $\det 8$ and 150 ms for $\det 16$. The illustrated timing diagram applies to a full SZ reloading cycle (occurring, e.g., every 10th cycle). For standard cycles involving only a replace from SZ, the reload operation column under the blue block consists solely of the black dashed box, and the ``LZ to SZ \textnormal{\&} {MZ}" operation is replaced by ``SZ to MZ", reducing the overall duration to $\sim 30~\text{ms}$. The rearrangement duration to shuttle atoms from the MZ to the RZ (or vice versa) is $2.9~\text{ms}$ when running the $\det 8$ code which uses half of the MZ sites, and $5.8~\text{ms}$ when running the $\det 16$ code which uses the entire MZ.}
    \label{Extended Data Fig:ext_reload_time}
\end{figure*}
\endgroup

\section{1Q gates}
Our system uses two forms of single-qubit gates: global and local.  The local gates have been described in previous work \cite{barnes2022assembly, 2024AtomQEC}, and use two pairs of crossed-acousto-optic deflectors to apply fully arbitrary gates to atoms in the register in a row-parallel manner.  For this work, we have added the ability to perform global 1Q gates, where the same gate is applied to all atoms in parallel.  This reduces the time overhead associated with parallelizing over rows, and is less sensitive to alignment-induced gate quality fluctuations associated with the tightly focused local addressing beams.  We use the global 1Q gates for spin-echo pulses associated with MCM, and the local 1Q gates for computational operations.  

% The composite $\pi$ pulses, including those deployed later within the MCM block, are realized via a global beam illuminating the entire atomic array. 
The global gates are driven with a single beam with two RF tones separated in frequency by the qubit transition frequency, with the polarization carefully tuned to both provide the two components needed to drive the Raman transition between qubit states and to balance the light shifts on both qubit states. Each component pulse features a rounded, flat-top profile with a single $\pi$-pulse duration of $\sim 65~\mu\text{s}$. For spin echo pulses, we use a five-pulse Knill-type configuration to mitigate amplitude and second-order off-resonance errors.  For the specific phase sequence, see Table I (corresponding to Fig. 5b) of Ref.~\cite{Jones2013Designing}.  We bound the error associated with the full five-pulse composite $\pi$ pulse to $6\times10^{-4}$ by fitting the decay in return-probability when we perform many rounds of interleaved composite and random operations, followed by a pre-computed return operation.  The bound includes the contribution from the two non-composite $\pi/2$ pulses used to achieve randomization in each round.  

Because we do not perform a site-by-site reference-frame calibration between the global and local gates, mixing the two kinds of operation in a circuit requires that even numbers of global echo pulses are applied between any pair of local gates.

\section{Twisted toric codes}

In this section we describe the codes used in our experiment, relating their structure to the standard toric code. The construction is summarized here for convenience; the general framework and higher-dimensional generalizations are in Ref.~\cite{aasen2025}.

\begin{figure*}
    \centering
        \includegraphics[width=2.0\columnwidth]{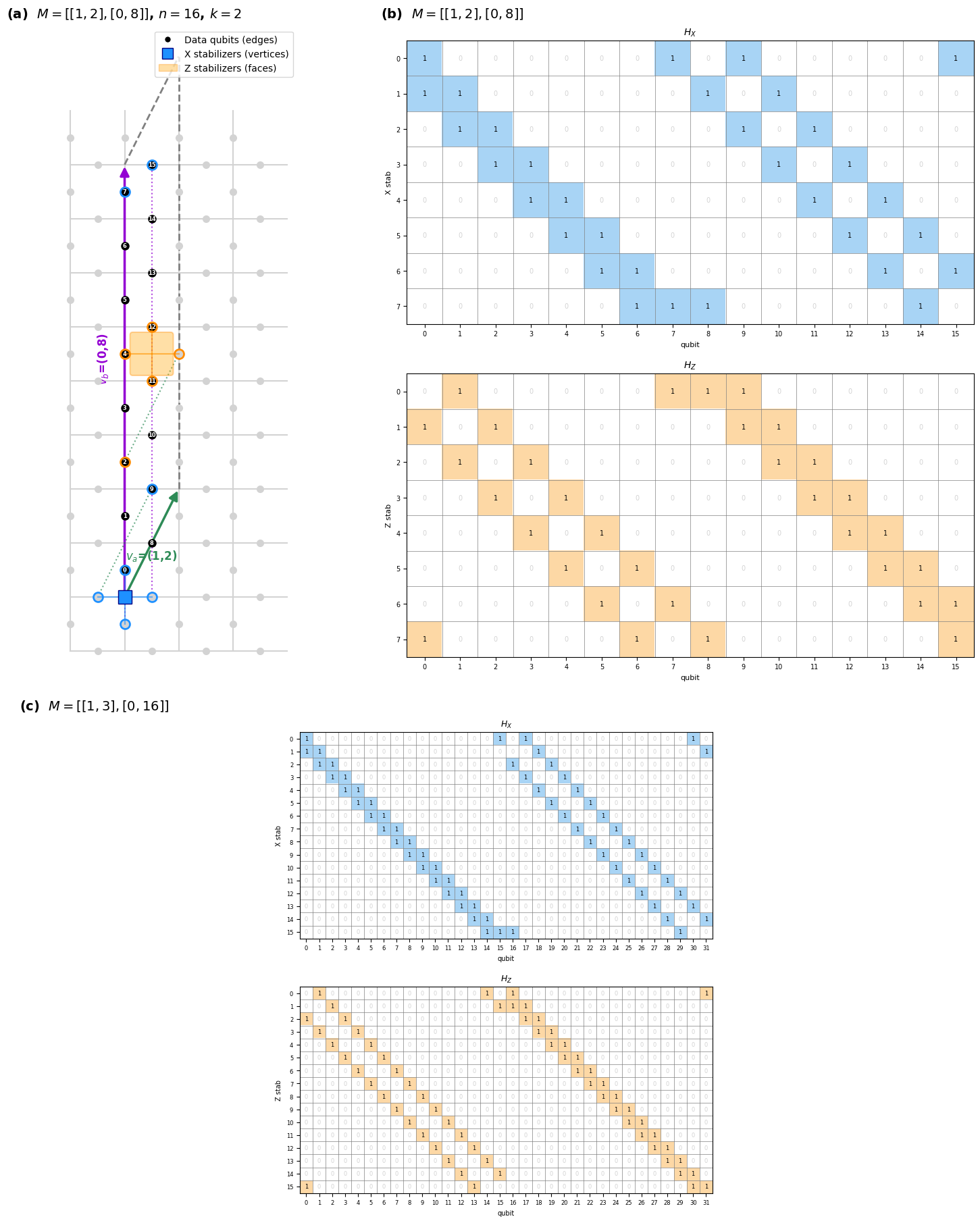}
    \caption{\textbf{(a)} Obtaining the twisted torus of the $\det 8$ code from the HNF. Starting with the square cellulation of $\mathbb{R}^2$ at integer coordinates, the rows of $M$ span a parallelogram: the fundamental domain of the torus, tiling $\mathbb{R}^2$ under translation by lattice vectors. Opposing edges of the fundamental domain are identified to yield the cellulated twisted torus. We can understand the geometry of the twisted stabilizer generators by drawing those of the standard toric code on the plane and translating any qubit outside of the fundamental domain into a qubit inside the fundamental domain via a unique lattice vector. \textbf{(b)} The stabilizer generators of the $\det 8$ code. \textbf{(c)} The stabilizer generators of the $\det 16$ code.
    }
    \label{fig:TTC}
\end{figure*}

The codes used in this work are two-dimensional toric codes on twisted tori, with the twist specified by a $2 \times 2$ integer matrix $M$ in Hermite normal form (HNF):
\begin{equation}
M = \begin{pmatrix} a & b \\ 0 & c \end{pmatrix}, \qquad a, c > 0, \quad 0 \leq b < c.
\end{equation}
The lattice $\Gamma \subset \mathbb{Z}^2$, defined as the points in the integer span of the rows of $M$, uniquely specifies the torus: two points of $\mathbb{R}^2$ represent the same point on the torus if and only if their difference is a lattice vector (see Extended Data Fig.~\ref{fig:TTC}).

\renewcommand{\figurename}{Extended Data Fig.}

%Obtaining the torus from the HNF. Starting with the square cellulation of $\mathbb{R}^2$ at integer coordinates, the rows of $M$ span a parallelogram serving as the fundamental domain of the torus, tiling $\mathbb{R}^2$ under translation by lattice vectors. Opposing edges of the fundamental domain are identified to yield the cellulated torus. \textbf{(a)} For diagonal $M$ ($b = 0$), the fundamental domain is the $a \times c$ rectangle, with opposite edges identified directly (left to right, top to bottom), recovering the standard untwisted toric code. \textbf{(b)} For $b > 0$, the fundamental domain is sheared into a parallelogram and the right edge is identified with the left after a vertical shift of $b$ (via the first generator $(a, b)$); the top is still identified with the bottom directly (via the second generator $(0, c)$). \textbf{(c)} [TBD: additional comment on twisted lattice].

Each integer-coordinate vertex in the fundamental domain contributes two outgoing edges---horizontal and vertical---carrying physical qubits, along with one plaquette. The domain contains $ac = \det M$ such vertices, giving $n = 2ac$ physical qubits. The standard local toric code prescription applies: an $X^{\otimes 4}$ stabilizer at each vertex and a $Z^{\otimes 4}$ stabilizer at each plaquette.

For the standard toric code ($b=0$, diagonal $M$), the minimum-weight logical operators are coordinate-aligned strings of weights $a$ and $c$, giving code parameters $[[2ac,\, 2, \,\min(a, c)]]$. Twisting distorts these minimum-weight representatives and the diagonal entries of $M$ no longer correspond to code s.
%--- a horizontal traversal of the torus must now combine $a$ horizontal steps with compensating vertical motion --- so the simple ``diagonal entries are the distances'' rule no longer applies. 
In general the code distance is the $\ell_1$ systole of $\Gamma$,
\begin{equation}
d = \min_{v \in \Gamma \setminus \{0\}} \|v\|_1.
\end{equation}
We use Gurobi to find minimum-weight representatives of logical operators of these codes. 
%which reduces to $\min(a, c)$ in the untwisted case but can differ in twisted lattices.

% As a result, the code distance---a single worst-case number over all logical operators---does not on its own describe how protection is distributed across the two logical qubits. The operationally relevant quantity is the per-qubit distance: for qubit $i$, $d_X^{(i)}$ is the minimum weight of any $X$-type logical that flips its $\bar Z$ observable, and $d_Z^{(i)}$ analogously. The basis generators $\bar X_i$, $\bar Z_i$ provide one logical operator in each equivalence class, but the \emph{product} logicals $\bar X_0 \bar X_1$ and $\bar Z_0 \bar Z_1$ also flip each individual qubit's complementary observable. In a twisted lattice, the product representative can be strictly lighter than either basis generator, in which case it sets the effective $d_X^{(i)}$ (resp.\ $d_Z^{(i)}$) below the basis weight. In the untwisted case the basis representatives are already the lightest available, so the distinction does not arise.

Using the qubit numberings shown in Extended Data Fig.~\ref{fig:TTC} (b) and (c), one can verify the following minimum weight representatives of the $X$ and $Z$ logical operators of the logical qubits probed by our experiments:
\begin{align}
\intertext{$\det 8$ code, $(d_X, d_Z) = (4, 3)$:}
\bar{X}_1 &= X_0 X_2 X_4 X_6, \\
\bar{Z}_1 &= Z_6 Z_7 Z_{15}, \\
\intertext{$\det 16$ code, $(d_X, d_Z) = (6, 4)$:}
\bar{X}_1 &= X_2 X_5 X_8 X_{11} X_{14} X_{16}, \\
\bar{Z}_1 &= Z_{10} Z_{11} Z_{12} Z_{27}.
\end{align}

For explicit calculations yielding the convention chosen for Extended Data Fig.~\ref{fig:TTC}, we enumerate the $ac$ vertices of the fundamental domain on a rectangular integer grid as $v_{i,j} = (i, j)$ for $i = 0,\, 1, \ldots,\, a-1$ and $j = 0,\, 1, \ldots,\, c-1$. Each vertex carries two qubits, on its outgoing horizontal and vertical edges: $q_\mathrm{h}(i, j)$ is the horizontal edge from $(i, j)$ to $(i+1, j)$, and $q_\mathrm{v}(i, j)$ is the vertical edge from $(i, j)$ to $(i, j+1)$. The boundary identifications follow from the lattice generators: $(a,\, j) \equiv (0,\, j - b \bmod c)$ via the first generator $(a, b)$, and $(i,\, c) \equiv (i,\, 0)$ via the second. 

\subsection{Swapped Syndrome Extraction Schedule}

For both $X$- and $Z$-type stabilizers, data qubits were scheduled in a E-S-N-W pattern as shown in Fig.~\ref{fig:circuit_diagram}.

\begingroup
\renewcommand{\figurename}{Extended Data Fig.}
\begin{figure*}[htbp]
    \centering
    \makebox[\textwidth][c]{\includegraphics[width=2\columnwidth]{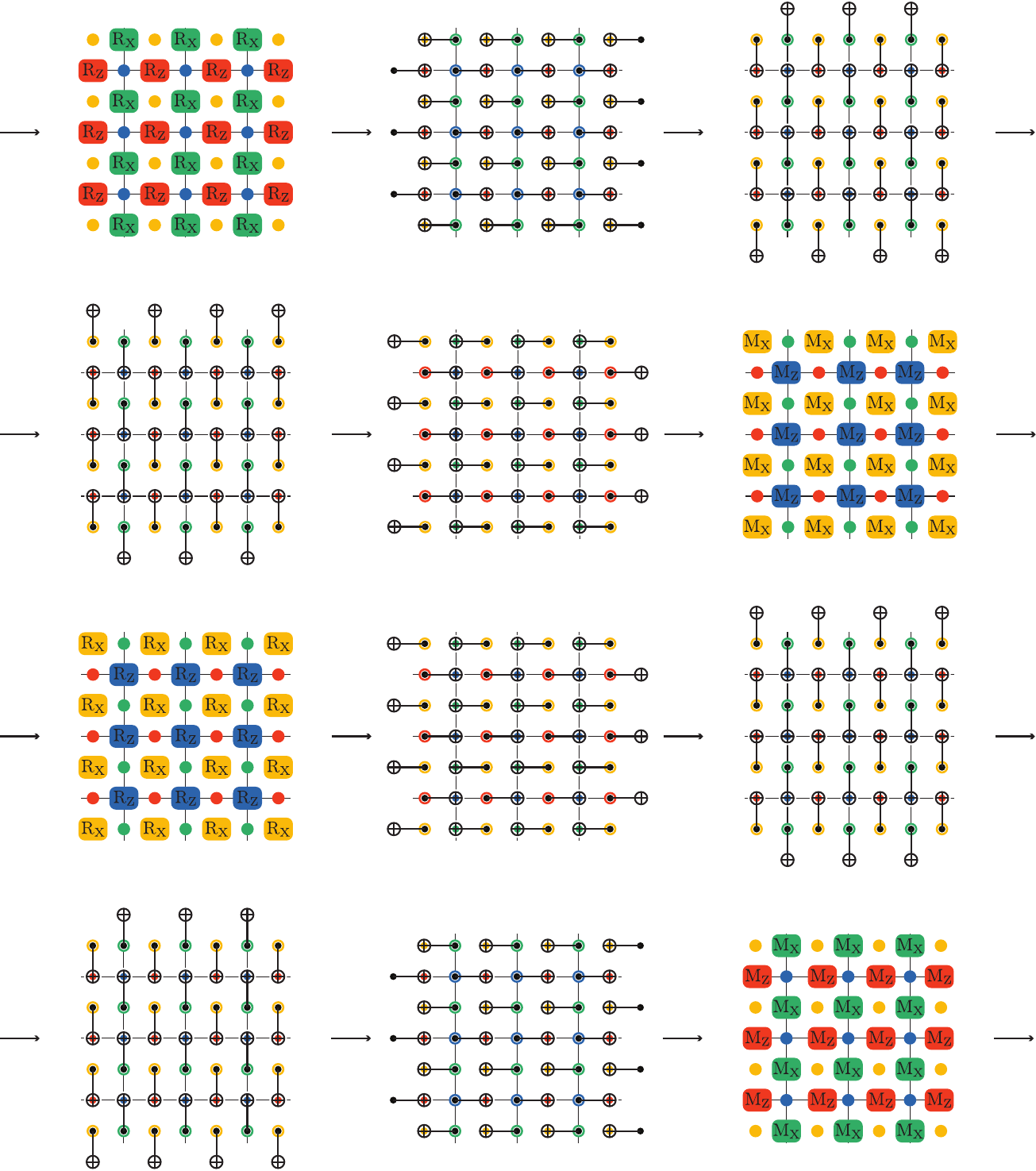}}
    
    \caption{Two rounds of loss-tolerant syndrome extraction. In each syndrome cycle, fresh qubits are initialized, four layers of gates perform syndrome extraction, and the former data qubits are measured. As a result, data and ancilla qubits exchange roles once per syndrome cycle so that every physical qubit is measured once every two syndrome cycles. These measurements enable qubit loss to be detected and corrected while simultaneously performing the usual stabilizer measurements.}
    \label{fig:circuit_diagram}
\end{figure*}
\endgroup

\section{Loss-aware decoding}

Although our circuits are constructed to identify and correct for atom loss, dealing with such losses during decoding presents a separate challenge. Although atom loss resembles an ideal erasure channel, we do not identify the exact temporal location of atom loss. Furthermore, we must deal with this delayed erasure in the presence of other errors that leave atoms within the qubit subspace \cite{baranes2025leveraging,liu2026_envelope}.

% To compute detection probabilities in which a lost atom was expected to contribute a bit value, we randomly assign values prior to computing detection probabilities. In an otherwise error-free circuit, a single loss before a measurement would then trigger 0 or 2 detectors with equal probability, leading to an average of 1 detection event per loss, versus weight-1 Pauli errors that contribute 2 detection events per error. The reduced sensitivity of detection probability to loss is consistent with reduced logical sensitivity to erasure error---although we note that atom loss is not immediately detected and is therefore not strictly an erasure error.

We follow the terminology of Stim, and formulate the decoding problem in terms of detector error models \cite{Gidney2021stimfaststabilizer,Derks2025}. For each possible error channel, a detector error model stores the prior likelihood that an error occurs, which detectors are flipped, and which logical observables are flipped. Using Stim it is straightforward to compute the set of detectors flipped by a given Pauli error at a specific circuit location.  
Our heuristic approach to decoding with atom loss is based on a simple model for Rydberg $CZ$ gates. 
If a qubit is lost prior to a $CZ$ gate, there is no Rydberg interaction between the lost qubit and its $CZ$ partner and we treat this absent interaction as a possible phase error on the partner qubit.

We observe that for individual atoms participating in a $CZ$ gate we expect $\langle Z\rangle = 0$. Ancilla qubits are initialized in $\vert +\rangle$ states (or other states in the $X-Y$ plane of the Bloch sphere) so that they are sensitive to the conditional phase of a $CZ$ gate. Similarly, single-qubit Pauli operators acting on data qubits have no expectation value. If we imagine such a qubit being lost before a $CZ$ gate and then reinserted after the $CZ$ gate, there is a relative phase ``missing'' between the $\vert 0 \rangle$ and $\vert 1 \rangle$ components of the lost-and-reinserted qubit's wavefunction.

We now outline the procedure for updating a detector error model in response to a lost qubit. For each lost qubit, we compute the likelihood $L_t$ that the qubit was lost when participating in a gate at timestep $t$. For each $CZ$ gate between the qubit's initialization and measurement, we introduce a new Pauli $Z$ error mechanism on the partner qubit at the circuit location immediately following the $CZ$ gate. We give this error mechanism a likelihood of $\epsilon=L_t/2$. The factor of $1/2$ reflects that the lost qubit had no expectation value for its single-qubit $Z$ operator prior to loss. Note that although an atom lost at timestep $t$ is also lost at timestep $t+1$, we ignore possible correlations between induced error mechanisms. In addition to the error mechanisms introduced on the partner qubits, we introduce a readout error with a likelihood of $50\%$ on the lost qubit immediately before readout while assigning the lost measurement a random bit-value.

In practice, we use Stim to pre-compute the set of updates associated with each possible observed loss. Thus, obtaining an updated detector error model can be efficiently computed for each bitstring. We note that this approach is compatible with any algorithmic decoder that accepts a detector error model (or some equivalent representation) as input. Examples of such decoders are Relay-BP and PyMatching \cite{Higgott2025sparseblossom,2025relaybp}.

Priors for the detector-error model are generated from the independently calibrated noise model detailed in Section \ref{sec:error_parameters} for data with atom reloads. For data without atom reloads, we use a noise model which reduces CZ loss and increases loss during ancilla reset after MCM.

\bigskip
\section{Error Parameters}\label{sec:error_parameters}

The simulated error sensitivity presented in Fig.~3(c) of the main text adopts a parameter set derived from independent measurements outlined below and summarized in Table~S1. The uncertainties quoted below represent the statistical fluctuations over time for most physical measurements; for parameters specifically derived from Gaussian decay channels, the error bars reflect the uncertainty in the fitting procedure. 

The single-site single-qubit (1Q) gate performance is benchmarked via standard Clifford randomized benchmarking (CRB)~\cite{Hashim2025Practical}, yielding an average error per-Clifford decay of $4.6(1.6) \times 10^{-4}$. This error is converted to a depolarizing Pauli stochastic error per native $SX$ gate of $p_{X,Y,Z}=1.15 \times 10^{-4}$ ~\cite{Nielsen2002Simple} (note we assume our virtual $Z$ rotations to be of unit fidelity). 

The two-qubit ($2\text{Q}$) errors are determined via interleaving $CZ$ gates within a reference sequence of two different random 1Q Cliffords applied to pairs of qubits (interleaved randomized benchmarking henceforth IRB)~\cite{Hashim2025Practical}. The reference sequence is generated such that the inverting 2Q Clifford for the interleaved sequence gives the total sequence \emph{depth} $CZ$ gates. The reference sequence includes the atom movement necessary to switch between 1Q (register) and 2Q gates (interaction zone), even though no 2Q gates are performed. The 2Q loss (here referring to at least one lost atom) is extracted from the additional loss in the interleaved sequence relative to the reference sequence, yielding $9.6(2.0) \times 10^{-3}$ per gate. This is mapped to a model where each atom is treated independently and equally likely to be lost. The error per $CZ$ gate post-selected on the survival of both atoms is extracted to be $4.3(1.9) \times 10^{-3}$. According to a theoretical model of our $CZ$ gates, this error is mapped to a 2Q Pauli stochastic error channel with probabilities $p_{IZ}=p_{ZI}=10p_{ZZ}$, where $p_{ZZ} = 1.9 \times 10^{-4}$.

We benchmark errors during movement and idling on a per-unit-time basis. Many of these are Gaussian processes, but our simulation pipeline is not currently able to use time-varying error probabilities to reproduce this behavior. Instead, we approximate Gaussian decay by matching the zeroth moment of a Gaussian function with time constant $\tau_{\text{decay}}$ over a maximum experiment timescale $t_{\text{total}}$ to an exponential decay with an inverse time constant $\frac{2t_{\text{total}}}{3\tau_{\text{decay}}^2}$. Here, $t_{\text{total}}\sim 400~\text{ms}$ is set corresponding to the duration of two full cycles of our $\det 16$ code, after which the qubits are reset via MCM. Note that this acts as an upper bound on errors in experiments with timescales shorter than $t_{\text{total}}$ due to the higher initial slope of the exponential function. These errors are marked with inverse time units.

With the grouping ``Idle stochastic", we account for three categories of single-qubit stochastic Pauli errors: qubits idling at RZ sites during circuits (\textit{idle Pauli}), qubits undergoing physical rearrangement (\textit{move Pauli}), and data qubits idling at RZ sites during mid-circuit measurement (\textit{MCM data qubit Pauli}). The \textit{idle Pauli} contribution is benchmarked using a Hahn echo sequence designed to mimic the circuit's built-in echo configuration. A Gaussian decay fit yields a coherence time of $T_2 = 7.2(0.6)~\text{s}$ from which we extract an exponential time constant of $5.1(0.9) \times 10^{-3}~\mathrm{s}^{-1}$. The \textit{move Pauli} error rate is assumed to be identical, as independent measurements reveal no discernible performance degradation during rearrangement. The \textit{MCM data qubit Pauli} error is parametrized per MCM cycle. Its Pauli $X$ component is characterized by averaging the spin-flip probabilities of the two qubit states over repeated MCM, yielding a stochastic Pauli $X$ error of $1.9(0.5) \times 10^{-3}$ per MCM cycle. The Pauli $Z$ component is benchmarked via a Ramsey sequence spanning an increasing number of MCM cycles; the resulting Ramsey contrast decay of $1.08(0.24) \times 10^{-2}$ is converted to a stochastic Pauli $Z$ error of $5.4(1.2) \times 10^{-3}$ per MCM cycle.

``Idle loss" mechanisms include qubit loss during rearrangement (\textit{move loss}), loss accrued while idling within static tweezers traps (\textit{idle loss}), and loss accrued for qubits within the static RZ traps during MCM (\textit{MCM data qubit loss}). The \textit{move loss} is measured by tracking atom survival probability with moves of increasing duration. This follows a a non-exponential Gaussian decay with time constant $0.9(.2)\text{s}$ yielding an exponential decay rate $2.7(1.1) \times 10^{-2}~\mathrm{s}^{-1}$ over the time of an average number of physical rearrangement moves per atom over two complete cycles. For the \textit{idle loss}, we apply our Gaussian approximation directly to the data from Fig.~4, yielding a loss rate of $2.3(0.2) \times 10^{-3}~\mathrm{s}^{-1}$. The \textit{MCM data qubit loss} is benchmarked with repeated MCM cycles and averaged over both qubit states to be $3.5(0.7) \times 10^{-3}$ per MCM cycle. Note that the data qubits incur an extra two-photon loss channel compared to the imaged ancilla qubits from the trapping light and residual MCM imaging light.

Finally, state preparation and measurement errors (SPAM) include contributions from readout errors and loss.  The readout error profiles differ slightly between regular and MCM images; however, they remain consistently $< 2.0(0.4) \times 10^{-3}$ per image and constitute only a minor fraction of the overall error budget. SPAM loss is divided into two distinct factors: initial array filling imperfections (\textit{initial load loss}, evaluated at $6.8(6.8) \times 10^{-3}$) and atom loss during reset after MCM (\textit{MCM reset loss}, evaluated at $9.5(4.6) \times 10^{-3}$ per MCM cycle. 

\begin{table*}[t]
    \centering
    \caption{\textbf{Summary of error parameters.} Summary of the benchmarked physical error model parameters - see text for more details on benchmarking experiments and mapping to simulation parameters. These parameters are used in "Sim A".}
    \label{tab:layout_parameters}
    
    \renewcommand{\arraystretch}{1.3} % Row padding keeps vertical lines continuous
    \setlength{\tabcolsep}{13pt}       % Kept at 13pt for optimal width
    
    % Vertical bars '|' between every centered 'c' column
    \begin{tabular}{c | c | c | c | c} 
        Category & Parameter & Units & Benchmarked & Calibration Experiment\\
        \noalign{\hrule height 0.8pt} 
        1Q stochastic & SX total Pauli & per gate & $3.45(1.) \times 10^{-4}$ & 1Q CRB \\ 
        \noalign{\hrule height 0.3pt} 
        2Q stochastic & CZ total Pauli & per gate & $4.03(1.8) \times 10^{-3}$ & IRB \\ 
        \noalign{\hrule height 0.3pt} 
        2Q loss       & CZ loss        & per gate & $9.6(2.0) \times 10^{-3}$ & IRB\\ 
        \noalign{\hrule height 0.3pt} 
        \raisebox{-2.3ex}{Idle stochastic} & idle Pauli $Z$ & $~\mathrm{s}^{-1}$ & $5.1(0.9) \times 10^{-3}$ & Hahn echo (Gaussian) \\ 
                                      & move Pauli $Z$ & $~\mathrm{s}^{-1}$ & $5.1(0.9) \times 10^{-3}$ & Hahn echo (Gaussian)          \\ 
                                      & MCM data qubit Pauli $X$ & per MCM & $1.9(0.5) \times 10^{-3}$ & MCM state decay\\ 
                                      & MCM data qubit Pauli $Z$ & per MCM & $5.4(1.2) \times 10^{-3}$ & MCM Ramsey contrast decay\\ 
        \noalign{\hrule height 0.3pt} 
        \raisebox{-1.5ex}{Idle loss}  & move loss      & $~\mathrm{s}^{-1}$ & $2.7(1.1) \times 10^{-2}$     & Repeated moves (Gaussian)             \\ 
                                      & idle loss      & $~\mathrm{s}^{-1}$ & $2.3(0.2) \times 10^{-3}$  & Trap lifetimes (Gaussian)              \\ 
                                      & MCM data qubit loss & per MCM & $3.5(0.7) \times 10^{-3}$ & MCM survival decay\\ 
        \noalign{\hrule height 0.3pt} 
        SPAM (non-loss)               & readout      & per image & $< 2.0(0.4) \times 10^{-3}$ & Prepare and measure \\ 
        \noalign{\hrule height 0.3pt} 
        \raisebox{-0.8ex}{SPAM (loss)} & initial load loss & per load & $6.8(6.8) \times 10^{-3}$ & Prepare and measure \\ 
                                      & MCM reset loss & per MCM & $9.5(4.6) \times 10^{-3}$ & MCM survival decay \\ 
    \end{tabular}
\end{table*}

Note that while some errors like Pauli errors and readout infidelities are universal to all detectors, the manifestations of different loss mechanisms on detector probability are dependent on the chronological order and our qubit role-swap schedule (see Extended Data Fig.~\ref{fig:loss_vs_det}). This role-swapping of data and ancilla qubits effectively symmetrizes the error channels in the bulk sequence; however, this symmetry is broken at the boundaries. Specifically, the 0th, 1st, and final detectors capture additional or incomplete errors.

\begingroup
\renewcommand{\figurename}{Extended Data Fig.}
\begin{figure*}
    \centering
    \includegraphics[width=2.0\columnwidth]{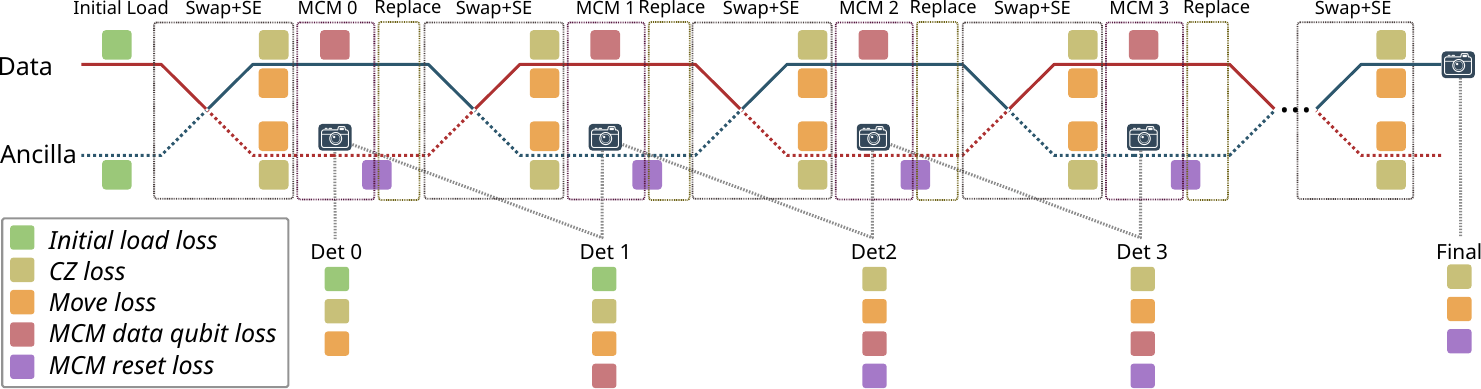}
    \caption{Impact of loss mechanisms on detector statistics. Data (solid line) and ancilla qubits (dashed line) are susceptible to distinct loss mechanisms throughout the circuit (see legend). Apart from the first and final cycles, each detector is sensitive to the losses within intervals over two consecutive MCMs.}
    \label{fig:loss_vs_det}
\end{figure*}
\endgroup
\section{Simulations}

\begingroup
\renewcommand{\figurename}{Extended Data Fig.}
\begin{figure*}
    \centering
    \includegraphics[width=2.0\columnwidth]{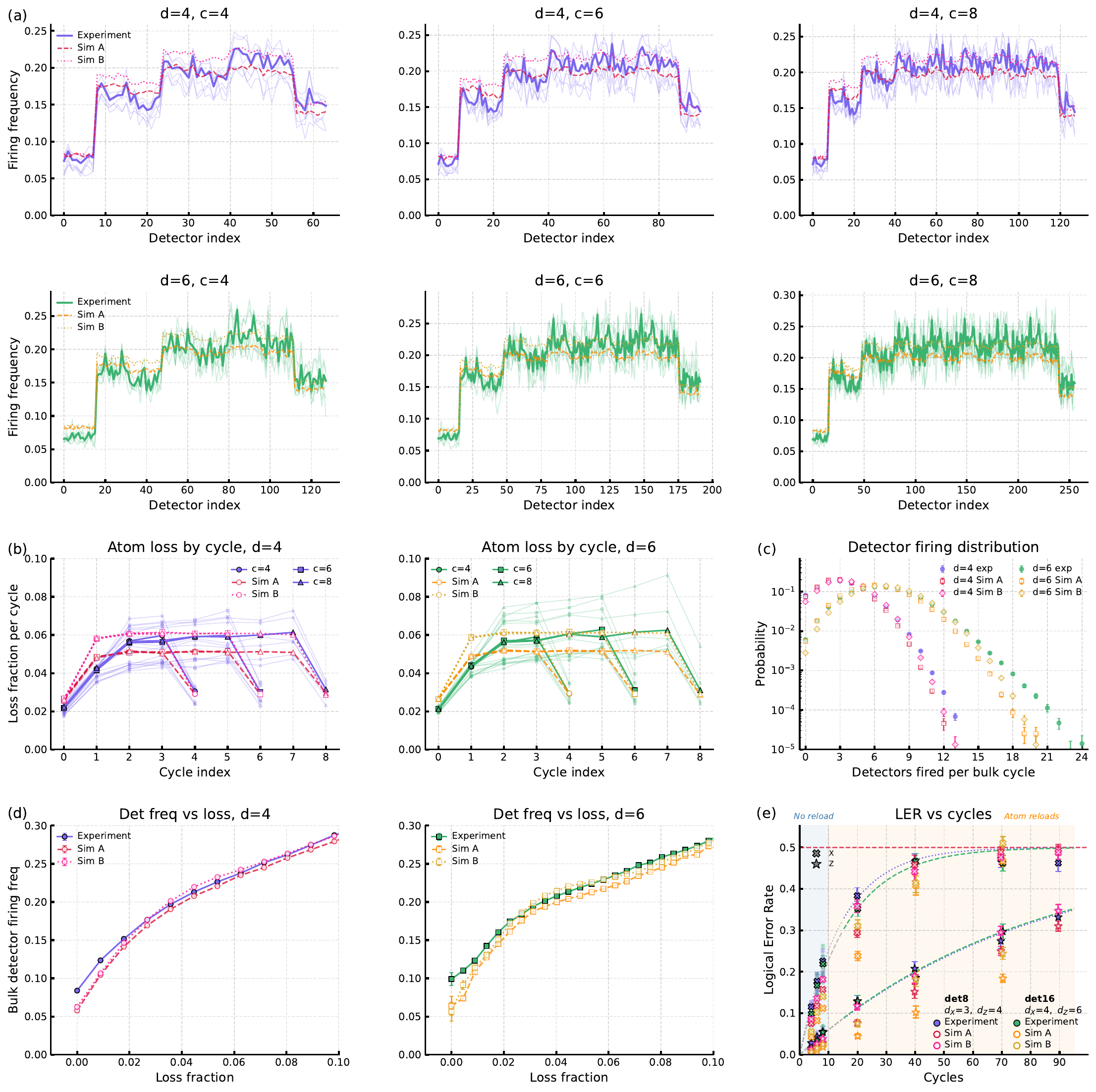}
    \caption{Comparison of experiment data to simulations. ``Sim A" is using an independently benchmarked error model while ``Sim B" phenomenologically adds 1\% to MCM data qubit loss. Semi-transparent lines are individual datasets, and solid lines are averaged over all of these datasets. (a) Average detector frequency by index for all data without reloads. Note $\det 8$ ($\det 16$) contains 16 (32) detectors per cycle, and the first and last cycles only contain Z detectors. For all other cycles, all X detectors are listed before all Z detectors. Note Sim A appears to underestimate late cycles, while SimB overestimates early cycles - we attribute this to slow atom heating during the circuit. (b) Atom loss by syndrome extraction cycle - a slow increase in loss across bulk rounds again appears which is not captured well by simulation, and we attribute this to slow atom heating. There is significant variation across datasets in both loss and detector frequency indicating slow drifts with time. (c) Bulk detector (detectors in rounds 2 to $c-1$ indexing from 0) firing distributions. The experimental data has a slightly longer tail than either simulation, indicating correlated errors which we are not currently accounting for (atom heating, coherent errors). (d) Bulk detector firing frequency as a function of measured ancilla loss fraction in that cycle. (e) Comparison of experimental and simulated logical error rates. Sim A underestimates the experimental logical error rates and predicts error suppression with distance with increasing cycles. The addition of extra MCM data qubit loss in Sim B recovers decreased error suppression with increasing cycles, although it still seems to underestimate errors at increased cycles, particularly for the $\det 16$ code.}
    \label{fig:sims}
\end{figure*}
\endgroup

Simulations were performed using a Clifford simulator including loss. Coherent errors and atom heating were not included due to the additional computational overhead required. Additionally, errors introduced by atom reloading were not considered in this simulation (we note that atom reloading causes an observable spike in detector frequencies around the reloading rounds, but it does not appear to have a substantial effect on logical error rate in our data).

We first compare simulations using a set of independently benchmarked parameters (Table \ref{tab:layout_parameters}, "Sim A") to several aspects of the experimental data. This model captures many features of the detector frequencies (Extended data Fig. \ref{fig:sims}(a)) including the slight elevation of X detectors over Z detectors as well as the firing frequency of early rounds; however, it underestimates "bulk cycle" detector frequencies. Turning to atom loss resolved by cycle (Extended Data Fig. \ref{fig:sims}(b)), we see a similar trend. Looking at the distribution of the number of detectors fired in a bulk cycle (Extended Data Fig. \ref{fig:sims} (c)), we see that the model captures the mean of the distribution reasonably, but underestimates the long tail which could be indicative of unaccounted for correlated errors \cite{clader2021impact}. This model underestimates logical error rates (Extended Data Fig. \ref{fig:sims} (e)), and further predicts an error suppression factor $\Lambda_Z>1$ at arbitrary cycles whereas our data appears to plateau to $\Lambda_Z\approx1$ at and above 8 cycles.

By phenomenologically adding an additional 1\% data qubit loss during MCM ("Sim B"), we better capture the detector frequencies and loss during bulk cycles (Extended Data Fig. \ref{fig:sims} (a) and (b)); however, we now overestimate these quantities in early cycles. This could be explained by slow atom heating during the circuit leading to increased loss with rounds. This model better predicts logical error rates in addition to the observed decrease in $\Lambda_Z$ with cycles. It still underestimates the tail of the bulk detector firing distribution (Extended Data Fig. \ref{fig:sims} (c)) as well as detector firing frequencies at low loss (Extended Data Fig. \ref{fig:sims} (d)), indicating that unaccounted for coherent errors likely also play a role in the observed logical error rates.

We note that the individual datasets shown in Extended Data Fig. \ref{fig:sims}(a) and (b) show significant variation in detector frequency and loss. The origins are currently unclear; however, possible explanations include drifting commensurability between the imaging cavity and tweezer arrays causing atom heating and loss as well as calibration timing.

We explore the contributions to our observed LERs by simulating these codes using the parameters in Table \ref{tab:layout_parameters}, and scaling each parameter individually to get partial gradients to the full LER. To do this, we simulated the LER for the full noise model, except one of the parameters is scaled down to zero ($\textrm{LER}_0$), and then we simulate the LER again with that parameter scaled by a factor f ($\textrm{LER}_f$). Then, to first order, we can approximate the gradient to the LER for each parameter with $\Delta_{\textrm{LER}} = \left( \textrm{LER}_f - \textrm{LER}_0 \right) / f$. The results of the partial gradients are categorized as in Table \ref{tab:layout_parameters}, and are depicted in Fig.~3(c) of the main text for the gradient range (0, 1) (i.e. $f=1$), with simulations using 100,000 shots. 

\section{Supplemental Data}
For completeness, we include here the data for the detector probability of the $\det 8$ toric code up to 90 cycles with SZ reloading (Extended Data Fig.~\ref{Extended Data Fig:det4x90}).

In Extended Data Fig.~\ref{fig:z_ler_by_date}, we break down the $Z$ logical error rates for the six individual datasets comprising Fig. 3 of the main text up to 8 cycles without reload. These itemized runs confirm that the observed logical error suppression of the $\det 16$ code over the $\det 8$ code remains persistent against experimental drift.

\begingroup
\renewcommand{\figurename}{Extended Data Fig.}
\begin{figure*}
    \centering
    \includegraphics[width=2.0\columnwidth]{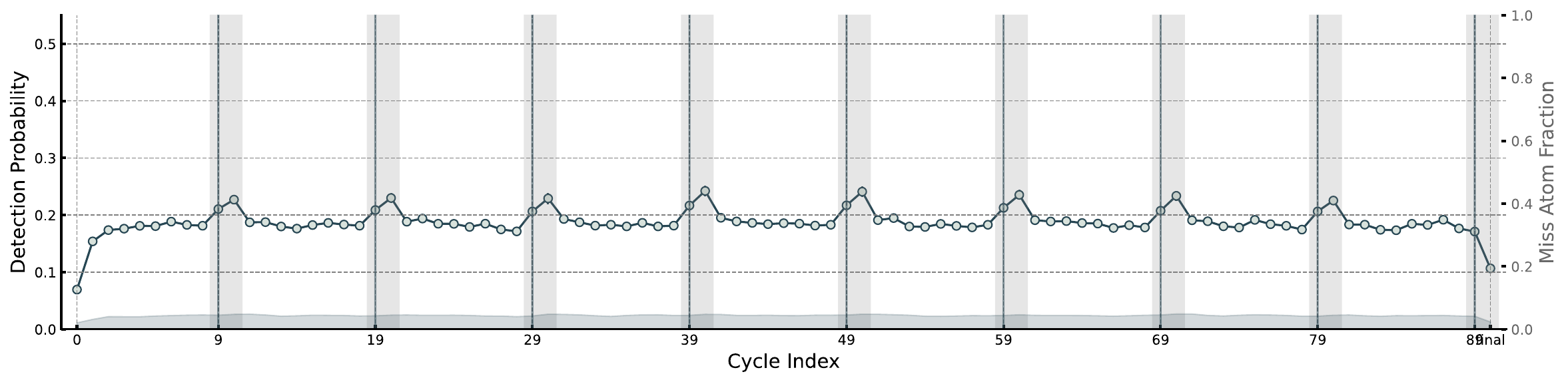}
    \caption{A $\det 8$ toric code detection probability over 90 consecutive syndrome extraction cycles with SZ reloading every 10th cycle. The missing atom fraction is shown in shades.}
    \label{Extended Data Fig:det4x90}
\end{figure*}
\endgroup

\begingroup
\renewcommand{\figurename}{Extended Data Fig.}
\begin{figure*}[htbp]
    \centering
    \makebox[\textwidth][c]{\includegraphics[width=1.3\columnwidth]{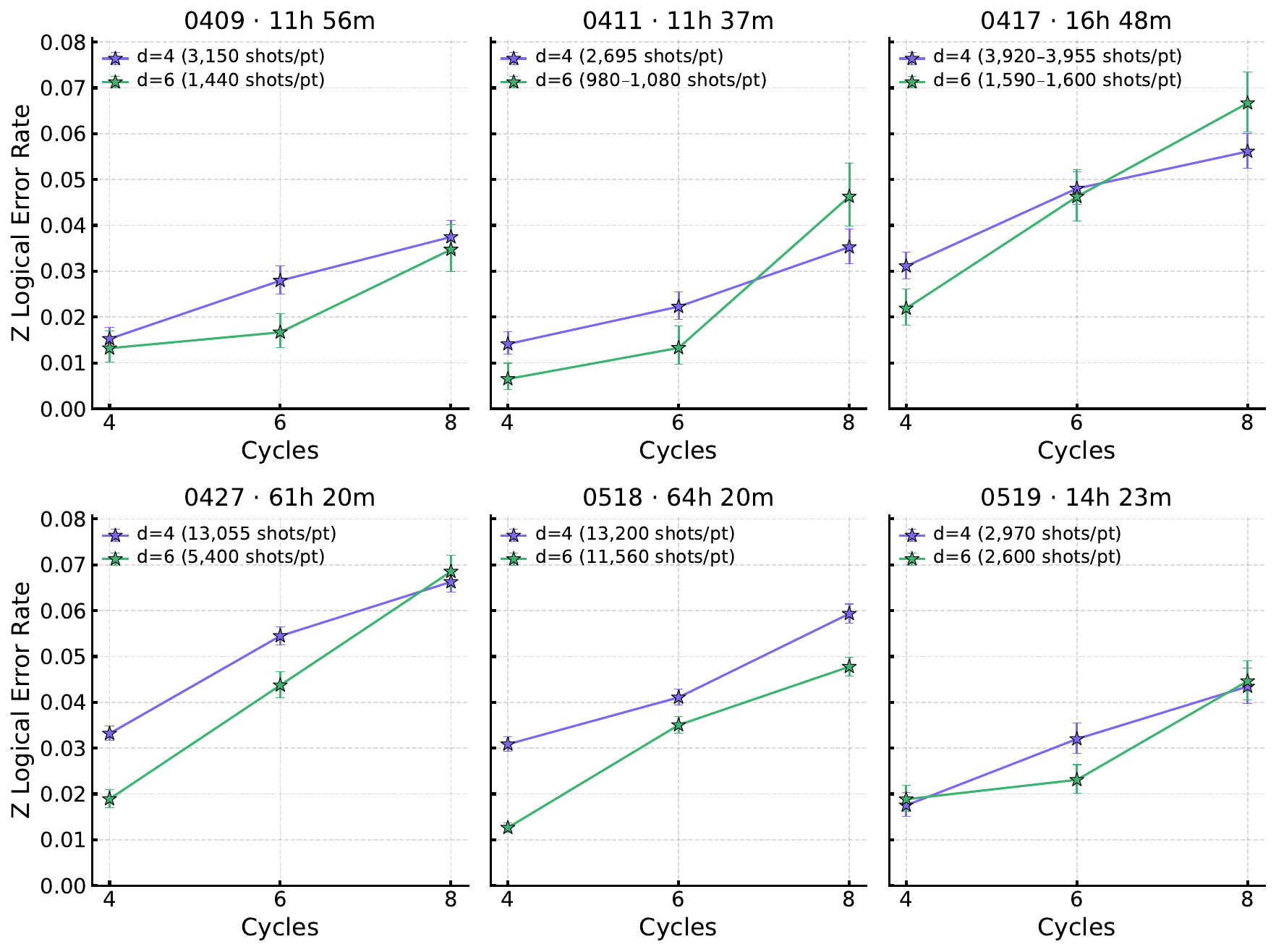}}
    
    \caption{$Z$ Logical error rates for the six individual datasets used in the demonstration of logical error suppression without atom reload. The number of shots per point and duration of each dataset is indicated in the sub-figures. While slow fluctuations of the experiment affect the logical error rates, we consistently see logical error rates for the $\det 16$ code ($d_X=6$) at or below that of the $\det 8$ code ($d_X=4$).}
    \label{fig:z_ler_by_date}
\end{figure*}
\endgroup

% \bibliography{bib} % let's deal with the citation later
% \end{document} 

\end{document}